\documentclass[11pt]{article}
\usepackage[margin=0.9in]{geometry}

\usepackage{multicol}
\usepackage{multirow}
\usepackage{tabularx}
\usepackage{booktabs}
\usepackage{colortbl}

\usepackage{graphicx} 
\usepackage{tikz}
\usetikzlibrary{shapes.geometric, arrows.meta, positioning, fit, backgrounds,decorations.text}
\usepackage{fontawesome5}
\usepackage{tikzpeople}

\tikzset{
  mathinner/.style={rectangle, rounded corners=3pt, draw=black!70, line width=.9pt,
                    fill=white, text width=5.90cm,
                    align=left, inner sep=5.2pt},
  codeinner/.style={rectangle, rounded corners=3pt, draw=deepgreen!90!black, line width=.9pt,
                    fill=white, text width=8.95cm,
                    align=left, inner sep=5.2pt},
  loopbox/.style={rectangle, rounded corners=3pt, draw=black!70, line width=.9pt,
                  fill=white, align=center, inner sep=5.2pt},
  arrow/.style={->, >=Stealth, line width=1.1pt, black!60},
  maparrow/.style={->, >=Stealth, line width=1.3pt, deepgreen!90!black},
  optarrow/.style={->, >=Stealth, line width=1.1pt, black!60, dashed},
    flow/.style={-{Stealth[length=3.1mm,width=2.45mm]}, line width=1.15pt, draw=black},
  innerflow/.style={-{Stealth[length=2.2mm,width=1.7mm]}, line width=.72pt, draw=black},
  axis/.style={-{Stealth[length=1.35mm,width=1.0mm]}, line width=.42pt, draw=black},
  edgelabel/.style={font=\small,align=center,inner sep=1pt}
}
\definecolor{pyrobg}{HTML}{FFF7E8}
\definecolor{teal}{HTML}{11BFB5}
\definecolor{orange}{HTML}{FF7900}
\definecolor{purple}{HTML}{8A2BE2}
\definecolor{workershirt}{HTML}{F28A2B}
\definecolor{deepgreen}{RGB}{124,176,102}
\definecolor{midgreen}{RGB}{143,198,135}
\definecolor{lightgreen}{RGB}{214,232,210}
\definecolor{optionalfill}{RGB}{232,243,229}
\colorlet{optionalbg}{optionalfill}

\newcommand{\puzzlepiece}[3]{%
  \begin{scope}[shift={#1},scale=#3]
    \path[fill=#2,draw=none]
      (-0.50,0.50) -- (-0.14,0.50)
      arc[start angle=180,end angle=0,radius=0.14]
      -- (0.50,0.50) -- (0.50,0.14)
      arc[start angle=270,end angle=90,radius=0.14]
      -- (0.50,-0.50) -- (0.14,-0.50)
      arc[start angle=0,end angle=180,radius=0.14]
      -- (-0.50,-0.50) -- (-0.50,-0.14)
      arc[start angle=-90,end angle=90,radius=0.14]
      -- cycle;
  \end{scope}%
}

\newcommand{\inlinepuzzle}[1]{%
  \tikz[baseline=-0.55ex,scale=.18]{%
    \path[fill=#1,draw=none]
      (-0.50,0.50) -- (-0.14,0.50)
      arc[start angle=180,end angle=0,radius=0.14]
      -- (0.50,0.50) -- (0.50,0.14)
      arc[start angle=270,end angle=90,radius=0.14]
      -- (0.50,-0.50) -- (0.14,-0.50)
      arc[start angle=0,end angle=180,radius=0.14]
      -- (-0.50,-0.50) -- (-0.50,-0.14)
      arc[start angle=-90,end angle=90,radius=0.14]
      -- cycle;}%
}

\def\alice{\tikz[baseline=-2ex]{
\node[alice,minimum size=0.5em] at (0,0) {};}
}

\def\bob{\tikz[baseline=-2ex]{
\node[bob,minimum size=0.5em] at (0,0) {};}
}

\newcommand{\githubsha}{2d30ce4de4e5745511e84a7230e6a70312af5bf3}
\newcommand{\github}[1]{\href{https://github.com/lucasgautheron/adaptive-tests/blob/\githubsha/#1}{\raisebox{-0.5ex}{\includegraphics[width=1em]{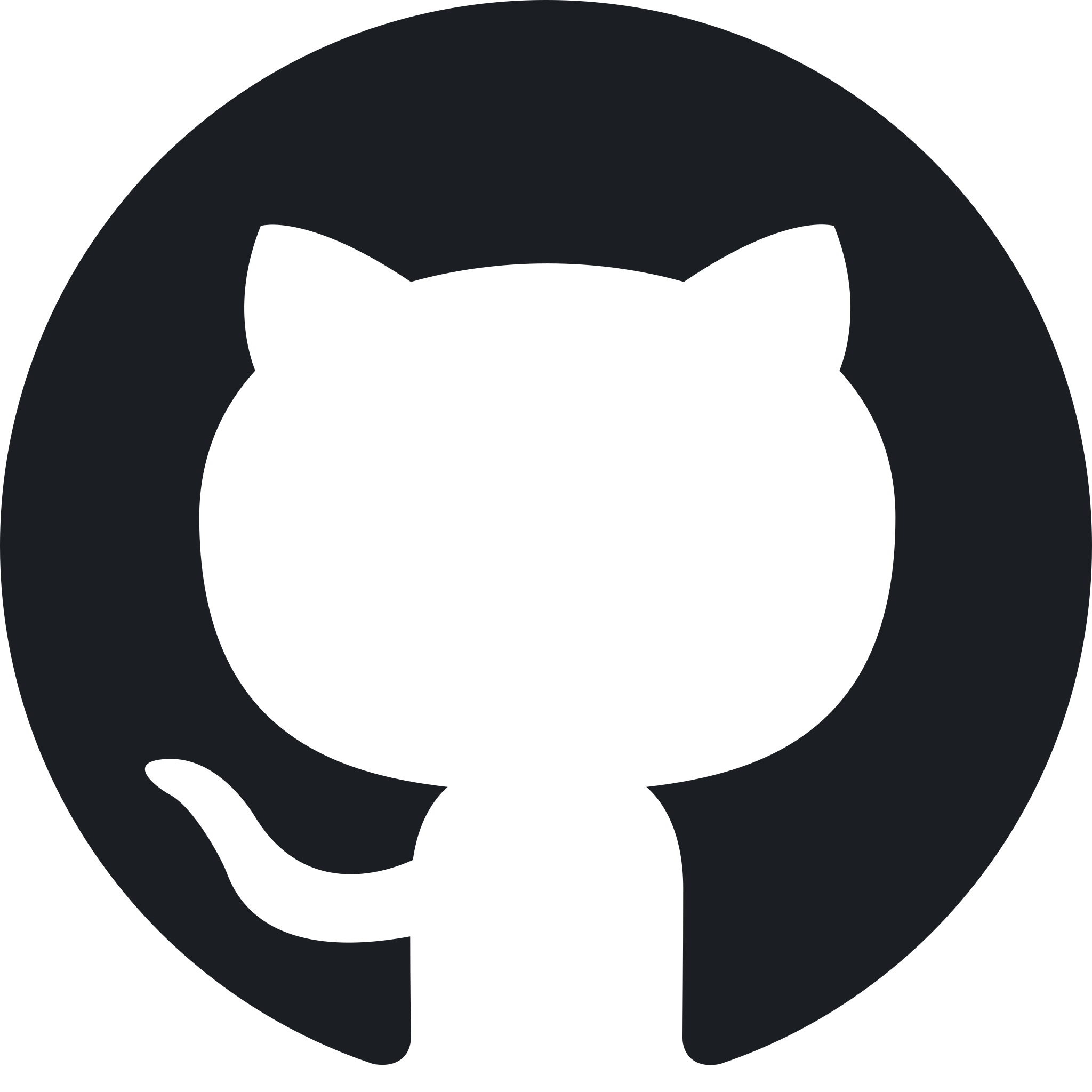}}}}

\usepackage{subcaption}
\usepackage{threeparttable}

\usepackage{amsmath}
\usepackage{amssymb}
\usepackage{bm}
\usepackage{bbm}
\DeclareMathOperator*{\argmax}{arg\,max}
\DeclareMathOperator*{\argmin}{arg\,min}
\usepackage{aligned-overset}

\usepackage[frozencache=true,cachedir=minted-cache]{minted}
\newcommand{\python}[1]{\mintinline[breaklines]{python}{#1}}

\usepackage{algorithm2e}

\usepackage{hyperref}

\usepackage{tablefootnote}
\usepackage[many]{tcolorbox}
\newtcolorbox[auto counter,number within=section]{mybox}[1][]{
  enhanced,
  fonttitle=\scshape,
  title={Infobox \thetcbcounter},
  #1
}

\usepackage[backend=biber, style=apa, sortcites=false,sorting=nyt,uniquelist=false,natbib=true,doi=true,isbn=false,eprint=false,pagetracker,ibidtracker=constrict, giveninits=true]{biblatex}
\usepackage[acronym,nonumberlist]{glossaries}
\makeglossaries
\newacronym{bad}{BAD}{Bayesian Adaptive Design}
\newacronym{ado}{ADO}{Adaptive Design Optimization}
\newacronym{eig}{EIG}{Expected Information Gain}
\newacronym{efe}{EFE}{Expected Free Energy}
\newacronym{dad}{DAD}{Deep Adaptive Design}

\newcommand{\codeurl}[1][]{\url{https://github.com/lucasgautheron/adaptive-tests/tree/\githubsha}}

\newcommand{\vect}[1][]{#1}

\title{Active Inference with People: a general approach to real-time adaptive experiments}
% ...

\author{Lucas Gautheron\textsuperscript{1}, Nori Jacoby\textsuperscript{2}, Peter Harrison\textsuperscript{3}\\
\textsuperscript{1}University of Missouri, \textsuperscript{2}Cornell University, \textsuperscript{3}University of Cambridge}
\date{}

\begin{document}

\maketitle

\begin{abstract}
Adaptive experiments optimize their design throughout data collection, which can bring substantial benefits compared to conventional experimental settings. Potential applications include, among others, computerized adaptive testing (when selecting informative tasks in ability measurements), adaptive treatment assignment (when searching for experimental conditions maximizing certain outcomes), and active learning (when choosing optimal training data for machine learning algorithms). However, implementing these techniques in real time poses substantial computational and technical challenges. In this paper, we introduce a practical and unified approach to real-time adaptive experiments that can encompass these scenarios across textual, visual, and audio tasks. Our strategy combines active inference, a Bayesian framework inspired by cognitive neuroscience, with Pyro, a probabilistic programming library, and PsyNet, a modular Python package for large-scale online behavioral experiments. Active inference provides a task-agnostic optimization objective and efficient inference strategies; probabilistic programming makes the computations practical, reducing implementation costs; and PsyNet makes the resulting procedure deployable with humans in real time across diverse behavioral paradigms. We illustrate this approach through two concrete examples: (1) an adaptive testing experiment estimating participants' ability by selecting optimal challenges, reducing the number of trials required by 30--40\%; and (2) an adaptive treatment assignment strategy that identifies the optimal treatment up to three times as accurately as a fixed design. We provide instructions to facilitate adoption of the workflow.
\end{abstract}



\tableofcontents

\newpage

\section{Introduction}


Adaptive experiments optimize their own design iteratively by leveraging the data already collected. Compared to traditional experimental designs, which rely on a fixed set of stimuli selected and assigned a priori, this general line of approach can provide significant benefits in diverse scenarios. For instance, in psychometrics, computerized adaptive testing seeks the most informative tasks to deliver to a subject, given the current estimate of the subject's ability and the difficulty of the tasks \citep{wainer2000computerized}. For example, if a participant failed easy tasks (like memorizing short sequences), it makes little sense to assign them a hard task (e.g., a longer sequence to memorize). Another relevant scenario is adaptive treatment assignment, in which accumulating evidence is used to  preferentially assign treatments that are more likely to achieve a desired outcome, with clear applications to clinical trials \citep{simon1977adaptive,Villar2015,ryan2020bayesian} or policy interventions \citep{kasy2021adaptive}. In psychology, adaptive experimentation may include the search for the conditions or stimuli that best serve  the objectives of an experiment. This includes, identifying stimuli that maximize differences in responses between different groups, as in computerized adaptive diagnosis \citep{Gibbons2016}. Finally, in machine learning, active learning seeks informative training data from humans by prioritizing samples for which the model is most uncertain \citep{houlsby2011bayesian}.

The above situations can be equally thought of as adaptive, iterated data-acquisition processes with humans in the loop. Namely, they describe experiments in which the stimuli presented to participants are informed or determined by the responses given in previous trials or by earlier participants. Such experiments require repeatedly updating what is known from the available data and deciding, under uncertainty, what data to acquire next. Bayesian frameworks naturally formalize these two operations and thus have the potential to provide general solutions across these problems. However, Bayesian methods can be too slow for real-time experiments or demand substantial implementation efforts that render them impractical. Therefore, their theoretical advantages remain difficult to leverage. Instead, these applications often rely on dedicated algorithms or rigid software tailored for particular use-cases\footnote{In computerized adaptive testing alone, optimization methods range from purely frequentist to fully Bayesian approaches and may include, to name a few: ``maximum Fisher information, maximum likelihood weighted information, maximum posterior weighted information, maximum expected information, minimum expected posterior variance, and maximum expected posterior weighted information'' \citep{Choi2009}. Implementations include R packages \citep{Magis2011,Chalmers2016}; or in adaptive psychometric testing, Mathematica scripts \citep{Watson2017} and JavaScript libraries \citep{Kuroki2022} designed for use within specific frameworks such as jsPsych \citep{deLeeuw2014} and PsychoJS \citep{Peirce2019}. These solutions are not only scattered across subfields, but also often limited to specific models and thus cannot benefit other scenarios in adaptive experiments. In particular, there is little overlap between these approaches and adaptive treatment assignment, a research area more often focused on bandit algorithms \citep{burtini2015survey}. A more ideal solution would remain agnostic about the nature of an experiment, enabling scientists to implement their own design with minimal effort.}. 

To address this gap, the present paper introduces a practical approach to real-time adaptive experiments that can flexibly encompass the above scenarios (Figure \ref{fig:solution}). This approach combines active inference, a broadly applicable Bayesian framework, with Pyro \citep{pyro_working_memory_tutorial}, a probabilistic programming library, and PsyNet \citep{harrison2020psynet}, a software package for large-scale online experiments. Active inference is a generalization of traditional \gls{bad} that can treat the above scenarios under the same mathematical apparatus. For the practical implementation, Pyro accepts arbitrary probabilistic response models and automates difficult computations via fast approximate inference algorithms. Finally, PsyNet can implement experiments across a wide range of paradigms and modalities (text, image, audio, video, \dots). Consequently, the proposed solution can facilitate the adoption of adaptive designs across the behavioral sciences.

We illustrate this strategy via two concrete examples: an adaptive testing experiment, and an adaptive treatment assignment setup.  To demonstrate the effectiveness of this approach, we conducted both experiments in simulations and with 200 human participants. Compared to static setups, our strategy achieves a 30--40\% reduction in the number of tests required in the first experiment, and a three-fold increase in accuracy for the second experiment, competing with or outperforming specialized algorithms.

The rest of the introduction proposes a quick review of Bayesian approaches to adaptive experimentation (\S\ref{section:bayesian}), and introduces active inference as a promising practical generalization (\S\ref{section:active_inference}). This is followed by a brief high-level description of PsyNet (\S\ref{section:psynet_intro}). Finally, our two concrete examples are introduced (\S\ref{section:experiments}).

\begin{figure}[!htbp]
    \centering
    \resizebox{0.75\textwidth}{!}{\input{flow}}
    \caption{Illustration of the proposed solution, in an experiment where the goal is to measure a latent participant trait $\theta_i$. \textbf{1.} PsyNet collects observations from the participants ($y_{1:t}$), and delegates computations to an appropriate probabilistic programming library such as Pyro. \textbf{2.} At every step, this library updates the posterior distribution of relevant latent parameters $\theta$ (which may encode a mix of individual-, group-, or population-level characteristics). \textbf{3.} Following active inference, it computes the \gls{efe} of each design (e.g., each test item, or each treatment), a measure that weighs expected information gains (EIG) with optional pragmatic costs or utilitarian objectives (U). \textbf{4.} The optimal design is selected. \textbf{5.} PsyNet then delivers the selected design to a participant in the appropriate modality. \textbf{6.} The participant submits their response through a specified channel.}
    \label{fig:solution}
\end{figure}

\subsection{\label{section:bayesian}Extant Bayesian approaches to adaptive experimentation}


In principle, Bayesian methods for adaptive experiments possess multiple advantages. First of all, they should be highly flexible, since they can in principle accommodate arbitrary tasks. Second, they are theoretically optimal, due to their ability to appropriately quantify uncertainty. This contrasts with heuristics handcrafted for specific settings, or even more principled frequentist strategies, neither of which usually attain optimality \citep[p.~3]{rainforth2023modernbayesianexperimentaldesign}\footnote{Regarding Fisher-information based strategies, \citet{rainforth2023modernbayesianexperimentaldesign} conclude that ``classical frequentist approaches to experimental design either require us to make undesirable approximations, ignore our uncertainty in $\theta$ despite this being our target for learning, and/or leverage aspects from the [Bayesian] framework itself.''}.

Bayesian approaches require two components. The first is the \textit{statistical model} that encodes the assumptions about the data-generating process. It is generally specified by a probability $p(y,\theta{\mid}d)$, where $y$ are the observable outcomes (e.g. the participants' responses); $\theta$ are unobservable latent parameters (say, the participants' ability, in computerized adaptive testing); and $d$ is the design (e.g. tests to administer, treatments to assign, or stimuli to classify). Adaptive experiments also require an \textit{objective function} that ranks the performance of each design, given the goals of an experiment. Among extant Bayesian approaches, we can distinguish two main categories, based on their objective: \gls{bad} and Bayesian decision theory.

\begin{table}[!h]
\centering
\begin{threeparttable}
\resizebox{\textwidth}{!}{%
\begin{tabular}{|l|c|c|c|}
\hline
\multicolumn{1}{|c|}{\textbf{Paradigm}\tnote{a}} & \textbf{\begin{tabular}[c]{@{}c@{}}Bayesian\\ adaptive design\\(BAD)\end{tabular}} & \textbf{\begin{tabular}[c]{@{}c@{}}Bayesian\\ decision theory\end{tabular}} & \textbf{Active inference} \\ \hline
\textbf{Objective} & \begin{tabular}[c]{@{}c@{}}Expected Information Gain:\tnote{b}\\ $\mathrm{EIG}(d) =H(\theta)-H(\theta|y,d)$\end{tabular} & \begin{tabular}[c]{@{}c@{}}Expected Utility:\\ $U(d) = \mathbb{E}_{p(y|d)}[r(y)]$\end{tabular} & \begin{tabular}[c]{@{}c@{}}Expected\\Free Energy:\tnote{c}\\$G(d) \simeq -[\mathrm{EIG}(d)+U(d)]$\end{tabular}\\ \hline
\textbf{Motivation} & \begin{tabular}[c]{@{}c@{}}Epistemic\\(maximizing knowledge)\end{tabular} & \begin{tabular}[c]{@{}c@{}}Pragmatic\\(seeking/avoiding \\specific outcomes)\end{tabular} & \begin{tabular}[c]{@{}c@{}}Epistemic\\\& pragmatic\end{tabular} \\ \hline
\textbf{Examples} & \begin{tabular}[c]{@{}c@{}}Computerized adaptive testing\\\& Active learning\end{tabular} & \begin{tabular}[c]{@{}c@{}}Adaptive\\ treatment assignment\end{tabular} & Any \\ \hline
\end{tabular}%
}
\begin{tablenotes}[flushleft]\footnotesize
  \item[a] Cf. the typology from \citet{sajid2022active}.
  \item[b] Following \citet{rainforth2023modernbayesianexperimentaldesign} we equate \gls{bad} to the maximization of the \gls{eig}, although the term is sometimes used in reference to different objective functions.
  \item[c] See, for instance, L3 of eq. (12) in \citet{Smith2022}.
\end{tablenotes}
\end{threeparttable}
\caption{\label{table:typology}Three Bayesian paradigms relevant to adaptive experiments\tnote{c}. Each approach starts with a generative model $p(y,\theta|d)$ and defines optimal designs (e.g., tasks, or treatments) according to an objective function. \textbf{Bayesian adaptive design} seeks the design $d$ that maximizes the expected information gain. \textbf{Bayesian decision theory} maximizes pragmatic/utilitarian goals by seeking designs achieving higher rewards. Finally, \textbf{active inference} naturally combines epistemic and utilitarian imperatives into a single objective $G$ called \gls{efe}. $G$ is roughly the sum of two contributions: the expected information gain (EIG), and an expected utility ($U$). In absence of preferences for specific outcomes ($U=0$), active inference reduces to \gls{bad} \citep{sajid2022active}.}
\end{table}



\paragraph*{Bayesian adaptive design} (\gls{bad}, sometimes referred to as \gls{ado}; \citealt{Cavagnaro2010,Myung2013}) seeks experimental setups that achieve the highest information gains, measured as the reduction in the entropy of target parameters $\theta$ (e.g., in adaptive testing, the individual abilities). This suggests a simple adaptive optimization strategy: choose the design with the highest Expected Information Gain (EIG) at every step of the experiment \citep{rainforth2023modernbayesianexperimentaldesign}. In the context of real-time adaptive testing, this strategy can effectively replace and outperform traditional algorithms, while applying to arbitrary response models. In fact, \gls{bad} applies to any experiment that aims to acquire information optimally, much beyond adaptive testing\footnote{Although it is not our focus, this paradigm can also extend to the selection of training data for machine learning algorithms \citep{houlsby2011bayesian}}. Yet, \gls{bad} can raise significant computational and implementation obstacles. First, this method requires tracking the posterior distribution of the relevant latent parameters $\theta$, but traditional MCMC methods can be too slow for models typically considered in adaptive testing without careful, manual tuning \citep{burkner2021bayesian}. Second, the estimation of the \gls{eig} itself is computationally difficult in most cases \citep{rainforth2023modernbayesianexperimentaldesign}.

\paragraph*{Bayesian decision theory}
The maximization of information gains is a purely epistemic objective; it is suitable when the goal of an experiment ``is not to reach decisions but rather to gain knowledge about the world'' \citep{Lindley1956}. However, certain experiments aim to uncover/leverage treatments and experimental conditions that maximize \textit{specific} outcomes or responses (see Table \ref{table:typology}). When such preferences are involved, Bayesian decision theory tends to be the paradigm of choice\footnote{\gls{bad} can also be seen as a Bayesian decision problem in which the reward is the information gain \citep{Bernardo1979}.}. Obvious applications include clinical trials \citep{simon1977adaptive,Villar2015,ryan2020bayesian} or policy choice \citep{kasy2021adaptive}. In this framework, the best design $d$ (e.g. the best treatment, policy, or experimental condition) is by definition the one with the highest expected utility $U(d)=\mathbb{E}_{p(y{\mid}d)}[r(y)]$, where $r(y)$ is the reward associated with outcome $y$. However, systematically delivering the treatment associated with the highest expected utility (i.e. payoff) at each iteration of the experiment -- a strategy known as myopic or greedy optimization-- would be a poor choice. Indeed, doing so generally fails to explore potentially superior alternatives
.  Explicitly forward-looking optimization considers entire sequences of tasks, which is superior in the long run but also computationally challenging \citep{Myung2013,rainforth2023modernbayesianexperimentaldesign}. Instead, implementations in adaptive experiments traditionally rely on policies that effectively enforce a certain level of exploration without making explicitly forward-looking computations \citep{burtini2015survey}, e.g., by occasionally trying seemingly inferior treatments\footnote{An example of such a rule is: ``choose the best known treatment with probability $1-\varepsilon$ (exploitation), or any other treatment with probability $\varepsilon$ (exploration)''. This is known as the $\varepsilon$-greedy method \citep[pp.~33--36]{sutton2015reinforcement}.}. However, it is not always clear which policy is appropriate, depending on, among other things, the relevant tradeoff between short-term and long-term rewards \citep{simon1977adaptive,kasy2021adaptive}\footnote{Adaptive treatment assignment requires a choice between short-term rewards (achieving high success \textit{during} the experiment) and long-term rewards (minimizing opportunity costs when rolling out the best performing treatment after the conclusion of the experiment). Adaptive experiments can aim for short-term rewards, long-term rewards, or any balance of the two, as well as other objectives (like achieving accurate rankings or simply identifying the best treatment), each warranting distinct policies in principle \citep{simon1977adaptive,kasy2021adaptive,burtini2015survey,horn2022comparison}.}. Moreover, traditional policies do not side-step difficult Bayesian computations, whose implementation can remain delicate.

\subsection{\label{section:active_inference}Active inference}

The present work introduces a third Bayesian approach based on active inference, a joint theory of perception and action in biological systems \citep{parr2022active}. This theory suggests solutions for implementing adaptive behavior in autonomous systems, artificial intelligence, digital twins \citep{lanillos2021active, Torzoni2026}, and as recently proposed, human-computer interactions \citep{MurraySmith2025}. Active inference casts both action selection and belief updating within a common variational framework. Agents seek actions that resolve the divergence between their desires and their expectations by minimizing a quantity called expected free energy (\gls{efe}, denoted $G$), while variational free energy provides the basis for updating beliefs about the world as observations accumulate. These two key features provide practical advantages in the context of adaptive experiments.

First, the \gls{efe} effectively combines the \gls{eig} and an optional expected utility contribution $U$ (Table \ref{table:typology}). This provides a heuristic for optimizing pragmatically motivated experiments in a myopic fashion: while the \gls{eig} term in $G$ implements a curiosity bias (i.e. a tendency for exploration), the utility term acts as a pragmatic bias for high-performing actions (i.e. a tendency for exploitation). Weighing these two terms can facilitate the discovery of effective treatments when future steps cannot be explicitly considered in the objective, offering a useful heuristic under limited time constraints\footnote{Naturally, myopic active inference is generally suboptimal compared to explicitly forward-looking optimization when evaluated against objectives such as the regret in bandit formulations of adaptive treatment assignment. Nevertheless, active inference can outperform traditional RL policies in certain configurations of bandit problems \citep{Markovi2021,Wakayama2023,Wakayama2025}; we provide further evidence of that on metrics and bandit algorithms commonly used in adaptive experimentation (Appendix \ref{appendix:active_inference_mab}). This is because the \gls{eig} term enables efficient exploration by leveraging contextual information (e.g., participant traits) that might affect the amount of information obtained by administering one versus another treatment.}. Importantly, epistemic and pragmatic objectives are not independent: actions that improve knowledge can alter future opportunities for reward, while reward-seeking actions determine what information becomes available. Active inference captures this interaction by optimizing these objectives jointly. Moreover, this strategy can also incorporate pragmatic \textit{costs}, even in experiments that only aim to maximize knowledge; in computerized adaptive testing, for instance, this can be used to progressively dismiss tests that often take too long to complete (a possibility demonstrated in Appendix \ref{appendix:slow}), making traditional Bayesian adaptive design cost-aware.

Second, active inference appeals to variational inference, an approximate posterior updating mechanism. Variational methods can achieve substantial performance gains both when tracking the posterior distribution of relevant parameters \citep{Blei2017,advi,svi} and when estimating the \gls{eig} \citep{foster2019variational}. Interestingly, Pyro can automatically perform these variational computations on arbitrary user-specified models, making real-time inference practical.

From a conceptual perspective, active inference encodes plausible heuristics in boundedly rational adaptive agents that can be leveraged in the computationally demanding case of real-time adaptive experimentation. From a practical perspective, it provides a minimal toolbox covering a wide array of scenarios in adaptive experimentation, and directly serves our goal to leverage opportunities for interactions between extant approaches.  Together with Pyro, it provides a flexible and often sufficiently fast computational framework for a wide range of adaptive experiments. To fully exploit this generality of purpose, however, it must be paired with an experimental framework capable of supporting an equally broad range of tasks and modalities. PsyNet provides this complementary flexibility.

\subsection{\label{section:psynet_intro}PsyNet} 

PsyNet provides a high-level Python framework for implementing online experiments with complex designs, including adaptive paradigms (Figure \ref{fig:psynet}). It builds on Dallinger to automate deployment, manage computing resources such as servers, and run experiments with integrated databases and monitoring tools. In addition, PsyNet can automatically connect with platforms such as Prolific, Amazon Mechanical Turk, or CINT to recruit participants and, if necessary, handle payments.

\begin{figure}
    \centering
    \resizebox{0.9\textwidth}{!}{\input{psynet}}
    \caption{\textbf{The infrastructure of PsyNet experiments}. \textbf{A. Software}. PsyNet experiments are implemented in modular, object-oriented Python code. \textbf{B.  Deployment.} Experiments are deployed on remote servers (self-hosted or cloud-based) and accessible to participants from their web browser. \textbf{C. Recruiting services.} Multiple participant-recruitment platforms are supported, including Prolific and MTurk.}
    \label{fig:psynet}
\end{figure}

Thanks to this automation, PsyNet enables very large-scale studies involving thousands of participants and numerous experimental conditions \citep{Marjieh2024}. Its modular architecture supports a wide variety of inputs, from standard HTML \citep{niedermann2024studying} to audio and video recordings \citep{AngladaTort2023,AngladaTort2022} and even Unity-based games \citep{Tchernichovski2023}. Finally, PsyNet's native support for defining dependencies between trials and participants makes it especially well-suited for a broad spectrum of adaptive experiments. For instance, PsyNet offers its own implementations of Markov chain Monte Carlo with people (MCMCP; \citealt{NIPS2007_89d4402d}) and Gibbs sampling with people (GSP; \citealt{harrison2020psynet}), which extend the traditional sampling algorithms to human input, enabling the efficient exploration of multi-dimensional perceptual and representational spaces in domains such as musical chords \citep{Marjieh2024}, prosody \citep{rijn21_interspeech}, voice characterization \citep{vanRijn2022}, and the study of visual aesthetics \citep{van2024using}\footnote{The goal of MCMCP/GSP is to obtain samples that faithfully represent the underlying high-dimensional distribution, allowing researchers to estimate its properties, such as its modes, variability, or differences between groups. MCMCP and GSP are constructed such that each trial corresponds to one iteration of an MCMC algorithm (classic MCMC or Gibbs sampling); only after many such trials can the distribution of interest be approximated. This differs fundamentally from the approach presented in the current paper, which aims to perform optimization and computation rather than distributional sampling.}.  
PsyNet also implements non-Bayesian adaptive algorithms like genetic algorithms with human feedback \citep{van2022bridging}.

On the technical side, PsyNet uses Python. This means it can easily interact with probabilistic programming libraries that can tackle difficult inference tasks efficiently. This would be much less straightforward in JavaScript-based experimental platforms such as Empirica \citep{Almaatouq2021}. Furthermore, owing to its object-oriented nature, the architecture of PsyNet is highly modular; this means that the users can make radical changes to their design (such as switching the modality of the task from text to audio, or switching from a static to an adaptive setup) with minimal changes to their code. This modularity also makes it straightforward to build upon the efforts of researchers from other fields. Finally, PsyNet can simulate entire experimental runs via custom response-generating functions, allowing researchers to test their adaptive experiment end-to-end before deployment.

\subsection{\label{section:experiments}Two illustrative examples}

We illustrate and demonstrate our workflow in two experiments covering two distinct scenarios (Table \ref{table:experiments}). In both experiments, we use trivia questions about the solar system and American history \citep{Dubourg2025,Mercier2026}. The goal of the first experiment is to infer the extent of participants' knowledge in a target domain as efficiently as possible. The goal of the second experiment is to find trivia questions that can effectively discriminate participants with and without college education. Despite their differences, we will show that these two setups can be  brought under the same computational framework and implemented with the same PsyNet code. 

\begin{table}[htbp] \centering \begin{tabular}{|p{3cm}|p{5cm}|p{6cm}|} \hline & \multicolumn{1}{c|}{\textbf{Experiment 1} (\S\ref{section:adaptive_testing})} & \multicolumn{1}{c|}{\textbf{Experiment 2} (\S\ref{section:adaptive_treatment})} \\ \hline \textbf{Paradigm} & Computerized adaptive testing & Adaptive treatment assignment \\ \hline \textbf{Goal} & Determine participants' knowledge of a topic & Find treatments most correlated with education-level \\ \hline \textbf{Budget} & Adaptive: stop delivering questions when information gains are $< \varepsilon$ & Fixed: five treatments per participant \\ \hline \multirow{4}{*}{\textbf{Objective}} & \multicolumn{2}{c|}{\textbf{Expected free energy G}} \\ \cline{2-3} & \multicolumn{1}{c|}{$G \simeq -EIG$} & \multicolumn{1}{c|}{$G \simeq -[EIG + U]$} \\ & \multicolumn{1}{c|}{Purely epistemic} & \multicolumn{1}{c|}{Pragmatic} \\ & \multicolumn{1}{c|}{(neutral)} & \multicolumn{1}{c|}{(targets/avoids specific outcomes)} \\ \hline \textbf{Result} & 30--40\% reduction in tests required & Up to 3$\times$ increase in accuracy (best treatment identification) \\ \hline \end{tabular} \caption{Comparison of experiments 1 and 2.} \label{table:experiments} \end{table}

 \ 

We now proceed as follows. In \S\ref{section:workflow}, we describe the workflow in more detail, starting with active inference with probabilistic programming (\S\ref{section:active_inference_pp}) and its integration in PsyNet (\S\ref{section:psynet}). In \S\ref{section:adaptive_testing} and \S\ref{section:adaptive_treatment}, we present Experiments 1 and 2 in turn, describing for each experiment its motivation, implementation, data, results, and discussion. Finally, in \S\ref{section:going-further}, we address potential limitations of the workflow and how they may be addressed in PsyNet. We discuss the possibility of real-time model evaluation, online learning, and deep adaptive design under this framework.

\section{\label{section:workflow}The workflow}

Below, we present the workflow used in the implementations of both Experiments 1 and 2.

\subsection{\label{section:active_inference_pp}Active inference with probabilistic programming}

The active inference workflow is summarised in Figure \ref{fig:aif_workflow}. At every step $t$ of the experiment, before choosing the next trial, we update the posterior distribution of relevant parameters based on available observations ($y_{1:t}$) via stochastic variational inference in general, or exact inference when possible. Then, we estimate the expected information gain (\gls{eig}) associated with each candidate design $d$\footnote{Enumerating all potential designs is feasible when the design-space is discrete, finite, and reasonably small. In other cases (e.g., when a stimulus is generated somewhere on a continuous scale, such as loudness, pitch, frequency, etc.), alternative approaches are nevertheless possible within this framework, using gradient-based optimization (see  \citealt[\S3.4]{rainforth2023modernbayesianexperimentaldesign}). In the present paper, we focus on discrete design spaces where all potential tasks can be enumerated.}. If relevant, we also estimate $U(d)$, the pragmatic/utilitarian contribution to the objective. Finally, we deliver the design that maximizes $\mathrm
{EIG}(d)+U(d)$, collect the participant's response, and repeat.

The implementation relies on Pyro \citep{bingham2019pyro}, a Python probabilistic programming library that provides off-the-shelf solutions for each step of this decision loop. With Pyro, users can flexibly specify their own statistical assumptions as Python code, and the computations will follow automatically without requiring manual mathematical derivations. Users need to specify (1) an observation model; (2) a simulation model; (3) often, a variational approximation for the posterior to speed up inference; (4) optionally, a stopping rule to decide adaptively when the experiment should end (as in Experiment 1); (5) for experiments that seek to avoid or maximize certain outcomes (as in Experiment 2), a function encoding preferences over outcomes.

\begin{figure}[!h]
    \centering
    \resizebox{0.85\textwidth}{!}{\newcommand{\code}[1]{\texttt{\footnotesize #1}}

\newtcolorbox{MathPanel}[1][]{%
    enhanced,
    colback = gray!5!white,
    colframe = black!70,
    colbacktitle = black!70,
    coltitle = white,
    fonttitle = \bfseries\large,
    title = {B. Statistical assumptions},
    halign title = flush center,
    boxrule = 1.2pt,
    arc = 2mm,
    outer arc = 2mm,
    left = 2mm, right = 2mm, top = 3mm, bottom = 3mm,
    remember as = math-frame,
    #1}

\newtcolorbox{PythonPanel}[1][]{%
    enhanced,
    colback = lightgreen,
    colframe = deepgreen!90!black,
    colbacktitle = deepgreen!90!black,
    coltitle = white,
    fonttitle = \bfseries\large,
    title = {C. Python declaration},
    halign title = flush center,
    boxrule = 1.2pt,
    arc = 2mm,
    outer arc = 2mm,
    left = 2mm, right = 2mm, top = 3mm, bottom = 3mm,
    remember as = python-frame,
    #1}

\newtcolorbox{DecisionPanel}[1][]{%
    enhanced,
    colback = gray!5!white,
    colframe = black!70,
    colbacktitle = black!70,
    coltitle = white,
    fonttitle = \bfseries\large,
    title = {A. Active inference decision loop},
    halign title = flush left,
    boxrule = 1.2pt,
    arc = 2mm,
    outer arc = 2mm,
    left = 3mm, right = 3mm, top = 3mm, bottom = 3mm,
    #1}

\begin{minipage}{18.4cm}

\begin{DecisionPanel}[width=18.4cm]
\centering
\begin{tikzpicture}[node distance=1.25mm]
  \node[loopbox, text width=0.95cm, minimum height=3.55cm] (data) {
    \textbf{Data}\\[5pt]
    $y_{1:t}$
  };
  \node[loopbox, text width=3.95cm, minimum height=3.55cm, right=of data, align=left] (infer) {
    \mbox{\textbf{Infer posterior}\;{\footnotesize $q_t(\theta)$}}\\[3pt]
    \code{svi = SVI(}\\
    \hspace*{0.20cm}\code{model, guide,}\\
    \hspace*{0.20cm}\code{Adam(...), Trace\_ELBO())}\\
    \code{svi.step(y, item\_ids)}
  };
  \node[loopbox, text width=3.05cm, minimum height=3.55cm, right=of infer, align=left] (eig) {
    \mbox{\textbf{Estimate EIG$(d)$}}\\[3pt]
    \code{eig = marginal\_eig(}\\
    \hspace*{0.20cm}\code{design\_model, designs,}\\
    \hspace*{0.20cm}\code{"y", ...)}
  };
  \node[loopbox, fill=optionalbg, text width=4.10cm, minimum height=3.55cm, right=of eig, align=left] (objective) {
    \textbf{Estimate $U(d)$}\\[4pt]
    \code{y\_s[d] = Predictive(}\\
    \hspace*{0.20cm}\code{design\_model, num\_samples=S)(d)["y"]}\\
    \code{U = torch.log(}\\
    \hspace*{0.20cm}\code{p\_star[y\_s.long()]).mean()}
  };
  \node[loopbox, text width=2.90cm, minimum height=3.55cm, right=of objective, align=left] (choose) {
    \textbf{Choose design}\\[5pt]
    \code{d\_hat = torch.argmax(eig + U)}
  };

  \draw[arrow] (data) -- (infer);
  \draw[arrow] (infer) -- (eig);
  \draw[arrow] (eig) -- (objective);
  \draw[arrow] (objective) -- (choose);

  \node[loopbox, text width=2.85cm, minimum height=1.35cm, below=3mm of choose] (observe) {
    Deliver $\hat{d}$; observe $y_{t+1}$
  };
  \draw[arrow] (choose.south) -- (observe.north);
  \coordinate (returnlow) at ([yshift=-5mm]observe.south);
  \coordinate (rr) at (observe.south |- returnlow);
  \coordinate (rl) at (data.south |- returnlow);
  \draw[arrow] (observe.south) -- (rr) -- (rl) -- (data.south);
  \node[font=\small, anchor=south] at ($(rl)!0.42!(rr)+(0,0.7mm)$) {append $y_{t+1}$ and repeat};
\end{tikzpicture}
\end{DecisionPanel}

\vspace{1mm}

\noindent
\begin{minipage}[t]{7.0cm}
\vspace{0pt}
\begin{MathPanel}[width=7.0cm]
\centering
\begin{tikzpicture}[remember picture, node distance=2.3mm]
  \node[mathinner] (m1) {
    \textbf{1. Observation model}\\[1pt]
    Priors and observation likelihood\\[2pt]
    \makebox[\linewidth][c]{$p(y_{1:t},\theta|d_{1:t})$}
  };
  \node[mathinner, below=of m1] (m2) {
    \textbf{2. Simulation model}\\[1pt]
    Outcome model for design $d$\\[2pt]
    \makebox[\linewidth][c]{$p(y_{t+1},\theta\mid d)\approx q_t(\theta)\,p(y_{t+1}\mid\theta,d)$}
  };
  \node[mathinner, fill=optionalbg, below=of m2] (m3) {
    \textbf{3. Variational approximations}\\[2pt]
    \makebox[\linewidth][c]{$q_{\phi}(\theta)\approx p(\theta\mid y_{1:t})$}
  };
  \node[mathinner, fill=optionalbg, below=of m3] (m4) {
    \textbf{4. Stopping rule}\\[1pt]
    Minimum information gain\\[2pt]
    \makebox[\linewidth][c]{$\mathrm{EIG}<\epsilon$}
  };
  \node[mathinner, fill=optionalbg, below=of m4] (m5) {
    \textbf{5. Outcome preferences}\\[1pt]
    Preference distribution over outcomes\\[2pt]
    \makebox[\linewidth][c]{$p^*(y)$}
  };
\end{tikzpicture}
\end{MathPanel}
\end{minipage}\hfill
\begin{minipage}[t]{10.8cm}
\vspace{0pt}
\begin{PythonPanel}[width=10.8cm]
\centering
\begin{tikzpicture}[remember picture, node distance=2.3mm]
  \node[codeinner] (c1) {
    \code{def model(y, item\_ids):}\\
    \hspace*{0.45cm}\code{ability = pyro.sample(}\\
    \hspace*{0.90cm}\code{"ability", dist.Normal(0., 1.))}\\
    \hspace*{0.45cm}\code{difficulty = pyro.sample(}\\
    \hspace*{0.90cm}\code{"difficulty", dist.Normal(0., 1.)}\\
    \hspace*{1.30cm}\code{.expand([n\_items]).to\_event(1))}\\
    \hspace*{0.45cm}\code{pyro.sample("y",}\\
    \hspace*{0.90cm}\code{dist.Bernoulli(logits=ability - difficulty[item\_ids]),}\\
    \hspace*{0.90cm}\code{obs=y)}
  };
  \node[codeinner, below=of c1] (c2) {
    \code{def design\_model(d):}\\
    \hspace*{0.45cm}\code{posterior = guide(y, item\_ids)}\\
    \hspace*{0.45cm}\code{ability = posterior["ability"]}\hfill{\scriptsize $\#\; q_t(\theta)$}\\
    \hspace*{0.45cm}\code{difficulty = posterior["difficulty"][d]}\hfill{\scriptsize $\#\; q_t(\delta_d)$}\\
    \hspace*{0.45cm}\code{pyro.sample("y",}\\
    \hspace*{0.90cm}\code{dist.Bernoulli(logits=ability - difficulty))}
  };
  \node[codeinner, fill=optionalbg, below=of c2] (c3) {
    \code{guide = AutoNormal(model)}
  };
\end{tikzpicture}
\end{PythonPanel}
\end{minipage}

\begin{tikzpicture}[remember picture, overlay]
  \draw[maparrow] (m1.east) -- (c1.west);
  \draw[maparrow] (m2.east) -- (c2.west);
  \draw[maparrow] (m3.east) -- (c3.west);
\end{tikzpicture}

\vspace{0.6mm}
\hfill
\begin{tikzpicture}
  \node[draw=black!60, rounded corners=2pt, fill=optionalbg, minimum width=0.55cm, minimum height=0.32cm] (legbox) {};
  \node[anchor=west, font=\small] at ([xshift=2mm]legbox.east) {Optional};
\end{tikzpicture}

\end{minipage}}
    \caption{A workflow for adaptive experiments driven by active inference using probabilistic programming. \textbf{A.} The experiment determines the optimal design at each iteration by inferring the current posterior $p_t(\theta)$, estimating the expected information gain and (if relevant) the expected utility of each candidate design $d$; and choosing the design that maximizes $\mathrm{EIG}(d)+U(d)$. Each of these steps can leverage pre-existing Python functionalities, side-stepping model-specific calculations.  \textbf{B.}~Researchers formulate several assumptions. (1) The model relating past observations $y_{1:t}$ to latent parameters $\theta$. (2) The predictive statistical model giving the probability of future observations given the latent parameters and a candidate experimental design (e.g. a potential test item or treatment). (3) A variational approximation $q_t(\theta)$ for the posterior $p(\theta{\mid}y_{1:t},d_{1:t})$ at time $t$, whose purpose is to achieve faster inference. In Pyro, this approximation is defined by a \textit{guide}. (4) An optional stopping rule, deciding dynamically when it is no longer worth administering tasks. (5) A preference distribution over outcomes, to focus on designs with desirable features (e.g., to identify high-performing treatments, or to avoid costly tasks). \textbf{C.}~Statistical decisions are declared as Python code via flexible probabilistic programming software such as Pyro.  }
    \label{fig:aif_workflow}
\end{figure}

\paragraph*{(1) Observation model}
The first ingredient is a statistical model for the observations collected so far:
\begin{equation}
p(y_{1:t},\theta\mid d_{1:t})=p(y_{1:t}\mid \theta,d_{1:t})p(\theta)
\end{equation}
where $d_{1:t}=(d_1,\ldots,d_t)$ denotes the sequence of tasks administered until iteration $t$, $y_{1:t}=(y_1,\ldots,y_t)$ the corresponding observed outcomes, and $\theta$ the latent quantities that govern those outcomes. For example, in computerized adaptive testing, ($d_t$) identifies the test item presented on trial ($t$), $y_t\in{0,1}$ records whether the participant answered it incorrectly or correctly, and the latent parameters include the participant's ability and the difficulties of the test items. The observation model specifies how likely the accumulated responses ($y_{1:t}$) are for different values of these latent quantities, given the items ($d_{1:t}$) that were administered; it therefore provides the basis for updating their posterior distribution as data accumulate. In our approach, researchers specify this model directly in Python using Pyro's probabilistic programming syntax (Figure~\ref{fig:aif_workflow}C), allowing for great flexibility.

\paragraph*{(2) Simulation model}

The workflow needs to predict the immediate outcome of each possible experimental ``design'' $d$. Formally, we need to specify $p(y_{t+1},\theta{\mid}d)$, the probability distribution for the future response $y_{t+1}$ under a candidate design $d$. This model must be supplied separately from the observation model because it does not at all involve prior observations. Instead, the simulation relies directly on the current posterior distribution of $\theta$ to simulate $y_{t+1}$, which is why this posterior is updated at every step of the decision loop.

For instance, in adaptive testing, the design will be the test item, and the simulation model will define the probability of success or failure given the current participant's own ability and the candidate item's difficulty. This probability depends on the current knowledge of $\theta$, i.e., its posterior distribution $p(\theta{\mid}y_{1:t},d_{1:t})$.

\paragraph*{(3) Variational approximation}

In the framework of active inference, agents continuously update their beliefs about $\theta$ via an approximate \textit{variational} inference scheme. This technique can considerably speed up inference compared to traditional MCMC methods, at the expense of lower statistical guarantees \citep{Blei2017,svi,advi}. In a nutshell, variational inference seeks an approximation $q_{\vect{\phi}}(\vect{\theta})$ to the posterior $p(\vect{\theta}{\mid}\vect{y}_{1:t},d_{1:t})$ by drawing from a family of distributions parameterized continuously by $\phi$, turning a sampling problem into an optimization task. Pyro offers a range of pre-defined choices named ``guides'', such as \texttt{AutoNormal}, which implements a very common normal approximation for the posteriors. We recommend starting from there before considering more sophisticated distributions if necessary. For instance, if the posteriors are weakly identified (e.g., the model can't tell if a test was easy, or the participant is skilled), \texttt{AutoMultivariateNormal} may be a better choice because it can better capture this form of posterior uncertainty. Other potential issues include strongly non-Gaussian posterior shapes or multimodality.\footnote{To name an example of strong departure from normality, take a 2PL model in adaptive testing with a very strongly sensitive test with difficulty $\delta$. If the participant passes the test $(y=+1)$, then $p(\theta<\delta|y)\simeq 0$; and if the prior $p(\theta)$ was normal, the posterior is a truncated normal. Multimodality is a more pathological situation that can arise in models with, say, mixture/latent classes models. In those cases, it is usually better to parameterize the model to avoid this issue.}.  Generally, the reliability of the inference procedure should be validated on synthetic samples drawn from the assumed statistical model. Appendix \S\ref{section:variational} introduces variational inference and its implementation in Pyro in more detail. In certain cases  (e.g., conjugate priors), the posterior distribution may admit closed-form expressions. For those, variational inference may be skipped altogether.

\paragraph*{(4) Stopping rule}

In computerized adaptive testing, it is common to stop delivering test items when some condition is met (e.g., the participant's ability is known with enough confidence) to minimize the number of tasks completed.
Although specifying a stopping rule may seem like an additional design requirement, we provide below some practical guidelines for selecting an appropriate criterion.
In active inference, a natural rule is to stop the experiment whenever $\max_d \text{EIG}(d)<\epsilon$, i.e., the expected information gain is lower than a minimum threshold. Indeed, assuming every design has a ``cost'' $\epsilon$, then $U(d)=-\epsilon$, and active inference prescribes doing nothing ($\emptyset$) if $\mathrm{EIG}(d)+U(d)<\mathrm{EIG}(\emptyset)+U(\emptyset)$, or equivalently, $\mathrm{EIG}(d)-\epsilon<0$. How to choose an appropriate numerical value for $\epsilon$? As a quick rule-of-thumb, assuming a single parameter $\theta$ with a Gaussian posterior $q(\theta)$, we have $\mathrm{EIG}\simeq \log \frac{\sigma_{t}}{\sigma_{t+1}} \simeq {\delta\sigma}/{\sigma_t}$, where $\sigma_t$ and $\sigma_{t+1}$ are the current and expected standard deviations of $q_t$ and $q_{t+1}$ respectively, and $\delta\sigma$ the expected reduction in standard deviation. Thus, the stopping rule $\mathrm{EIG}<\epsilon$ approximately requires
$\delta\sigma/\sigma_t < \epsilon$,
so continuing requires $\delta\sigma/\sigma_t \geq 10\%$ for $\epsilon=0.10$. What constitutes an acceptable threshold then depends on the cost of each task, but also on the pool of tasks at hand, and thus must be left to the discretion of the researcher. There are, however, hard limits on the range of appropriate values: for a binary outcome for instance, a single observation will contribute at most 1 bit of information, or $\simeq 0.693$ nat, which means $\epsilon$ must necessarily be strictly lower than 0.693. Appendix \ref{appendix:eig_2pl} compares the \gls{eig} with the Fisher information for a standard model in computerized adaptive testing. This can help researchers more familiar with the latter metric select an appropriate threshold for the \gls{eig} in this particular domain.

However, requiring that $\mathrm{EIG}>\epsilon$ can be too strict when optimizing the expected information gain greedily. In this case, it neglects the fact that while the next trial alone may provide limited information, a short sequence of trials may provide significant cumulative information gains. Thus, another strategy could be to deliver trials until the entropy of the parameters of interest (e.g. $H_{t}(\theta)$) goes below some threshold $\eta$. Under the same assumptions as in the above derivation of the quick rule-of-thumb, this gives $\sigma_{t}< \frac{\exp(\eta)}{\sqrt{2\pi e}}$; i.e., a minimum absolute uncertainty condition, similar to standard-error based stopping rules in computerized adaptive testing. This equivalence can be used to choose a suitable value for $\eta$. Owing to their model-agnostic information-theoretic framing, both criteria have the advantage of being broadly applicable (for instance, they extend much beyond computerized adaptive testing, and they are well-suited for knowledge maximization over multiple parameters at a time).

\paragraph*{(5) Preferences over outcomes}

Experiments may be driven by pragmatic goals in addition to information-seeking. In active inference, such goals are incorporated by weighing the \gls{eig} against a utilitarian term $U(d)\equiv \mathbb{E}_{q(y{\mid}d)}\left[\log p^{\ast}\right]$, where $p^{\ast}(y)$ is a probability distribution encoding preferences over possible outcomes.\footnote{Active inference casts optimal decision-making as Bayesian inference under a biased generative model, where the agent's preferences are encoded as priors $p^{\ast}(y)$ on $y$, which may be interpreted as a ``bias'' of the underlying generative model towards specific outcomes \citep{tschantz2020reinforcement}. This is not restrictive, since any reward function $r$ in an expected utility $U(d)\equiv \mathbb{E}_{q(y{\mid}d)}[r(y)]$ can be cast as a prior by setting $p^{\ast}\propto\exp(r)$. The probabilistic formulation also makes the \gls{eig} and expected utility commensurable, as both are measured in units of information.} This makes it possible to incorporate both desirable outcomes and pragmatic costs into the selection of experimental designs. For example:

\begin{itemize}
\item \textit{Avoiding costly tasks}. $U(d)$ can penalize designs associated with undesirable costs, such as tasks that take too long to complete. Appendix \ref{appendix:slow} illustrates this possibility using response time as a cost.

\item \textit{Identifying optimal treatments}. $U(d)$ can favor treatments or experimental conditions expected to produce desirable outcomes. Combining this term with the \gls{eig} balances \textit{exploitation} of treatments currently believed to perform well against \textit{exploration} of uncertain alternatives. We illustrate this application in Experiment 2.

\end{itemize}

\subsection{\label{section:psynet}PsyNet integration}

\begin{figure}[!htb]
    \centering
    \resizebox{\textwidth}{!}{\input{input}}
    \caption{\label{fig:architecture}\textbf{Components of a Bayesian adaptive experiment in PsyNet}. The timeline organizes the structure of the experiment. The trial maker delivers trials (i.e. tasks) to participants. Trials (i.e. individual tasks) are based on stimuli supplied by nodes, each node being a trivia question in our example. Each trial is associated with a specific participant-answer pair. The user interface displays the stimulus and collects the answer. Finally, an optimization module, called by the trial maker, performs the computational work and determines an optimal node at every trial. The code for each module can be accessed via the corresponding icon (\raisebox{-0.5ex}{\includegraphics[width=1em]{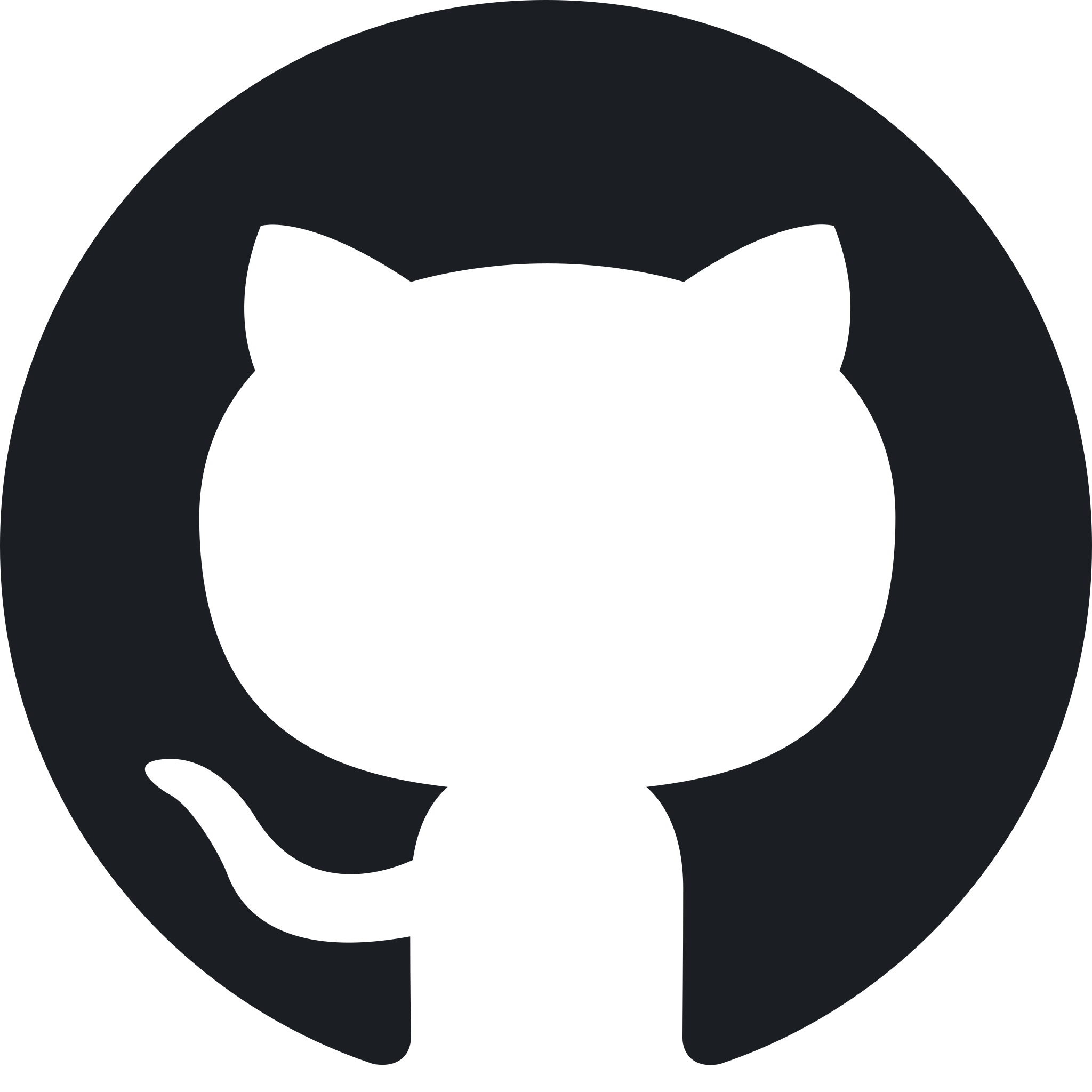}}).}
\end{figure}

The architecture of the PsyNet implementation of Experiments 1 and 2 is shown in Figure \ref{fig:architecture}. The crucial building blocks of any PsyNet experiment are the timeline; the trial makers; the trials; and the stimuli. The stimulus is typically stored in nodes which represent the stimuli state.
PsyNet is a highly modular object-oriented framework, such that each of these blocks can either directly import built-in implementations, or extend these implementations with additional/custom logic when necessary. The computational aspect of the Bayesian optimization process will be implemented in a separate class. This provides two benefits: (i) the optimization procedure can easily be tested in isolation and (ii) it will be very easy to change the optimization strategy or to switch back to a static design. As a result, we were able to use the same PsyNet code for both Experiments 1 and 2.

In what follows, we describe these basic components and their interactions. 
The entire code can be found on our GitHub repository \github{}\footnote{\codeurl} , in which both experiments were combined. Deploying PsyNet experiments online requires additional steps that are extensively described in the official documentation of the package\footnote{\url{https://www.psynet.dev/}}.The documentation also provides detailed descriptions of additional PsyNet components and a searchable reference for further information. In this tutorial, we focus on the design of the experiment itself.

\paragraph*{\label{paragraph:timeline}The timeline}

The first component of any PsyNet experiment is the \textit{timeline}
It describes the sequence of steps presented to the participant in an experiment. Timelines typically start by collecting consent and relevant demographic information. They include one or more ``trial makers'', whose role is to deliver tasks to the participants.

\paragraph*{\label{paragraph:trial_makers}Trial makers} In PsyNet, trial makers provide the logic through which ``trials'' (i.e. which tasks, or in our specific example, which trivia questions) are administered to participants. Among other things, the trial maker will be responsible for the implementation of the special logic through which optimal questions will be selected adaptively. To this end, we endow our custom trial maker with an \python{optimizer_class} attribute specifying which module will implement the computation of the optimal design at every iteration of the experiment. By making this class parameterizable, we make it easier to switch between potentially radically different optimization strategies while leaving the PsyNet modules untouched. This also enables the module to be tested in isolation from the rest of the experiment.  Trial makers instantiate trials of the appropriate type. 

\paragraph*{\label{paragraph:trials}Trials}

A trial is a single step of the experiment; it is characterized by a task, a participant, and the participant's response (see Figure \ref{fig:architecture}). Each trial is associated with a stimulus, called a \textit{node}: in our case, nodes are trivia questions
. 

\paragraph*{\label{paragraph:node}Nodes}

In PsyNet, nodes are persistent objects used to organize trials. A node stores the information from which its associated trials are constructed; in a static experiment, a node therefore typically corresponds to a reusable stimulus. In our experiment, each node represents one trivia question, such as ``What is the name of the first man that stepped on the Moon?''\footnote{Nodes are managed internally through networks, an abstraction that PsyNet also uses for more complex paradigms in which the experimental state can evolve across trials. In complex experiments with evolving stimuli, nodes are embedded within a network that represents their history. This network contains one or more nodes representing previous data states. Here we use PsyNet's static trial machinery: each question is represented by a single \python{StaticNode} in a non-growing, one-node network.}. We construct the nodes from a CSV file via \python{load_nodes()}, storing in each node's definition both the question and the list of admitted answers.

\paragraph*{\label{paragraph:interface}Interface}

In the \python{show_trial} method of this class, we use a \python{modular_page} to display both the question (\python{Prompt}) and the input (\python{Control}) through which the participants submit their answer. For example, textual inputs can use \python{TextControl}. 
PsyNet natively supports many types of \textit{prompts} (stimuli or inputs), including images (\python{ImagePrompt}), audio (\python{AudioPrompt}), videos (\python{VideoPrompt}), colors (\python{ColorPrompt}) and interactive canvas (\python{GraphicPrompt}). It also accepts diverse types of \textit{controls} (responses or outputs), including traditional HTML forms (dropdown menus, radio buttons and checkboxes, etc.),  participant recordings (audio, or video, cf. \python{AudioRecordControl} and \python{VideoRecordControl}), color sliders, drawings, etc. PsyNet supports custom HTML/JS frontends for tasks that require custom interfaces \footnote{ See the documentation section on ``Writing custom frontends.''}.
.

\paragraph*{Adaptive trial makers}

In order to make our design adaptive, we change the default logic through which PsyNet assigns tasks to participants by overriding the trial maker \python{find_nodes()} method \github{experiment.py\#L826-L843}. As part of this effort, we implemented a custom \python{prior_data()} method \github{experiment.py\#L759-L823} (not native to PsyNet) in the trial maker itself. We need this function because Bayesian optimization strategies leverage prior data to determine an optimal design; and these data have to be passed to the optimizer's  \python{get_optimal_node()} method, where the Bayesian computation takes place. This function is also responsible for updating the posteriors (via variational free energy minimization or exact computation), always reflecting the state of knowledge given all available data. 

\section{\label{section:adaptive_testing}Experiment 1: adaptive testing}

\subsection{Introduction}

We begin by implementing adaptive testing in a setup that assesses the participants' level of knowledge in a target domain using challenges of varying but a priori unknown difficulty. We emphasize the latter point, since many methods in computerized adaptive testing are limited to calibrated tests \citep{Fink2025}. We strive to retrieve as much information as possible about the participant's ability from as few trials as possible. Minimizing the number of trials offers several advantages: longer experiments can increase attentional fatigue; in online settings, they also raise the likelihood of participants dropping out before the end. More importantly, since participants are compensated based on the experiment’s duration, reducing the number of trials per participant allows us to recruit more participants for the same budget. 
As we will show, compared to a static setup delivering every item, the Bayesian optimization approach reduces the number of trials by 30--40\% without degrading the estimates of the participants' ability. To facilitate comparison with previous literature and begin with the simpler case, in this experiment, we limit ourselves to the \gls{eig} term of the expected free energy, such that active inference is identical to traditional \gls{bad}. 

\subsection{\label{section:experiments_aif}Methods}

In the first experiment, we aim to measure each participant's knowledge from as few trivia questions as possible, and we start without a calibrated item bank. Responses can be either correct ($y=1$) or incorrect ($y=0$). For the observation model (1), we use the traditional one-parameter logistic item-response model, where the probability of a correct answer depends on the participant's latent ability $\theta_i$ and a candidate-item's difficulty $\delta_d$:

\begin{equation}
    \label{eq:item_response}
    p(y_{i,d}=1{\mid}\theta_i,\delta_d) = \dfrac{1}{1+\exp{[-(\theta_i-\delta_d)]}}
\end{equation}

\noindent using priors $\theta \sim \mathcal{N}(0, 2), \delta \sim \mathcal{N}(b, 1), b \sim \mathcal{N}(0, 1)$, with $b$ the average question difficulty\footnote{A 2-parameter logistic model is a natural alternative, but more complex models can be poorly identified when the experiment starts with no prior data. When items are already calibrated, or when informative priors are available, richer models can be considered.}. The simulation model relies on the same assumption: for the current participant $i$, and for a candidate item $d$, we draw $(\theta_i,\delta_d)$ from the current posterior and simulate $y_{t+1,d}\sim\operatorname{Bernoulli}[\operatorname{logit}^{-1}(\theta_i-\delta_d)]$. 

The joint posterior over $(\theta,\delta,b)$ has no closed form, so the variational approximation is a mean-field family of independent Gaussians, i.e.: $p(\theta,\delta,b{\mid}y_{1:t},d_{1:t})\simeq q(\theta)q(\delta)q(b)$, where each $q$ is a normal distribution (see Pyro's \texttt{AutoNormal} guide). 
For the stopping rule, after each trial, we stop if $\max_d\operatorname{EIG}(d)<\epsilon=0.04$. 

Finally, there is no preferred outcome in this case, and the preference prior is flat, $\log p^{\ast}(y)=\mathrm{const.}$ Therefore, $U(d)=0$ and $G(d)\simeq-\operatorname{EIG}(d)$. Optimal questions are chosen by expected information gain, as in traditional \gls{bad}: Experiment~1 is the purely epistemic limit of active inference.

For the \gls{eig} computation, we used Pyro's \texttt{marginal\_eig()} implementation with 400 optimization steps and 400 samples per optimization step, followed by 10\,000 samples for the final Monte Carlo estimate (see Appendix \S\ref{section:variational_eig} for implementation details and tuning considerations). Readers interested in the details of the implementation can consult our code ~\github{experiment.py\#L285-L487}\footnote{\codeurl}. We also recommend Pyro's tutorial on adaptive experiments and its focused example (\citealt{pyro_working_memory_tutorial}; also \citealt{Myung2013} for additional conceptual background).

\subsection{\label{section:testing_data}Data}

We deployed the experiment in real time and successfully collected data from 200 participants recruited via Prolific, and in compliance with approved Max Planck Society Ethics Council protocols (2021-42). Additionally, we simulated the experiment on real human data from an ``oracle'' dataset from \citealt{Dubourg2025}, in which a large number of participants completed the full set of questions. This allows us to compare counterfactual strategies on the same individuals. Using PsyNet’s simulation functionality, we simulated participants whose responses to each selected item were drawn from the corresponding responses of individual participants in the oracle dataset. 

\subsection{\label{section:testing_results}Results and evaluation}

\begin{figure}[!htp]
    \centering
    \begin{subfigure}[t]{0.475\textwidth}
    \centering
    \includegraphics[width=0.95\linewidth]{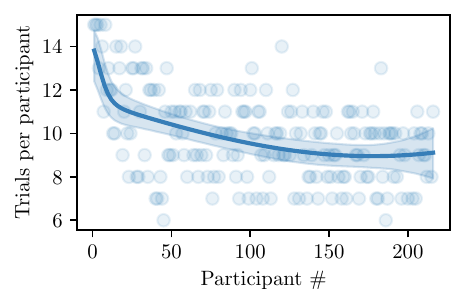}
    \caption{\textbf{Quantity of trials delivered per participant in the adaptive design.} Initially, almost all trials are administered to the participants, since doing so provides significant information gains about the questions' difficulty $\vect{\delta}$. Progressively, $\vect{\delta}$ is known more precisely, and the average number of trials decreases and converges.}
    \label{fig:trials_per_participant}
    \end{subfigure}\hfill\begin{subfigure}[t]{0.475\textwidth}
    \centering
    \includegraphics[width=0.95\linewidth]{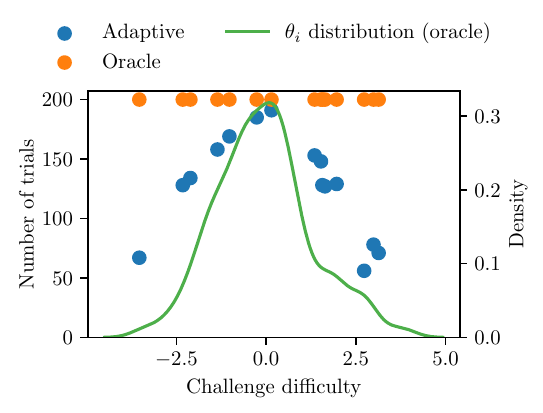}
    \caption{\textbf{Frequency of challenges delivered as a function of their difficulty $(\ast)$}. In the oracle, every challenge is presented to all of the 200 participants. In the adaptive setup, challenges that are either very easy (left) or hard (right) are administered much less frequently. Finally, the green curve represents the distribution of the abilities $\theta_i$.}
    \label{fig:challenges_distribution}
    \end{subfigure}
    \begin{subfigure}[t]{0.475\textwidth}
    \centering
    \includegraphics[width=0.95\linewidth]{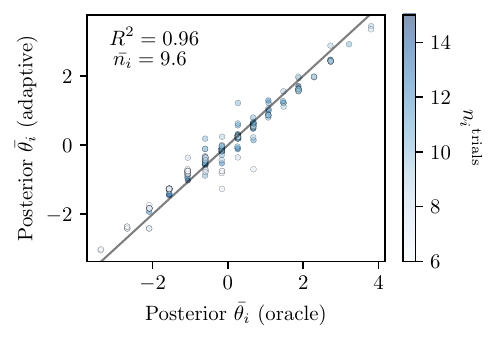}
    \caption{\textbf{Accuracy of the adaptive design $(\ast)$}. The x-axis represents the ``oracle'' estimate of each participant ability $\theta_i$, given the entire data (15 trials per participant). The y-axis represents the posterior estimate of $\theta_i$ in the adaptive design (9.6 trials per participant on average). Despite far fewer trials, the adaptive setup predicts the oracle estimates very accurately.}
    \label{fig:theta_estimate}
    \end{subfigure}\hfill\begin{subfigure}[t]{0.475\textwidth}
    \centering
    \includegraphics[width=0.95\linewidth]{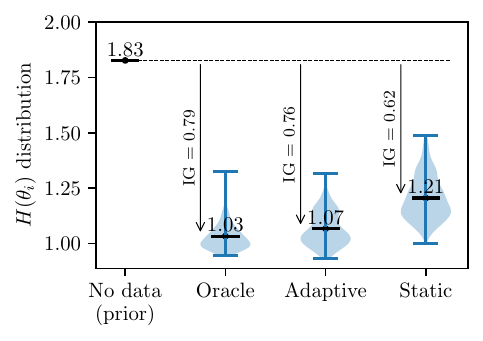}
    \caption{\textbf{Information gain (IG) across setups $(\ast)$}. The information gained through the experiment is the difference between the prior entropy of  $\theta_i$ (in the absence of data) and the posterior entropy of the ability $H(\theta_i)$ after the data collection. On average, the adaptive setup recovers almost as much information as the oracle. A static design, where questions are assigned randomly, with the same average number of trials as in the adaptive setup (9.6/participant), achieves lower information gain. }
    \label{fig:entropy}
    \end{subfigure}
    \caption{\label{fig:comparison}\textbf{Comparison of the adaptive and static setups on experimental data}. $(\ast):$ indicates results derived from counterfactual simulations using the human oracle dataset.}
\end{figure}

The adaptive design delivered an average of 9.9 trials per participant in the live deployment and 9.6 trials per participant in the simulation based on human data. In contrast to the oracle dataset, in which all 15 items were delivered to every participant, this amounts to a $\sim$ 30--40\%  reduction in the number of trials. This number varies substantially between individuals (see Figure \ref{fig:trials_per_participant}). The number of trials delivered is higher for the first participants, when the difficulty of the items is not known very precisely. The adaptive strategy quickly learns to optimize and reduce the number of trials. Figure \ref{fig:challenges_distribution} shows that the adaptive design delivers the easiest and hardest tasks less often. Indeed, such tasks are less often informative, because they are either trivial or impossible for most of the population. 

Is the reduction in tests delivered detrimental to the accuracy of the participants' ability estimate? To address this question, we contrast the estimates from the adaptive design with the oracle estimates (Figure \ref{fig:comparison}). The individual abilities $\theta_i$ measured in the adaptive setup correlate very strongly with the oracle estimate ($R^2=0.96$)\footnote{For this evaluation and the following, we used full Hamiltonian Monte Carlo sampling without approximation schemes to reconstruct the posterior abilities for each data collection strategy. For the evaluation of the entropy from the posterior samples, we use a Gaussian approximation, which gives $H=\frac{1}{2}\log (2\pi e\sigma^2)$ where $\sigma$ is the standard deviation of the samples. }. This relationship holds even for extreme abilities, which are associated with smaller numbers of trials.

This metric only assesses the quality of point-estimates. We further evaluated whether the posterior \textit{distributions} of $\theta_i$ were much more uncertain in the adaptive design than in the oracle by estimating the posterior entropy $H(\theta_i{\mid}\vect{y})$ for every participant
. The total information gained about each participant from the experiment \citep{Lindley1956} is directly given by the difference $\text{IG} = H(\theta_i)-H(\theta_i{\mid}\vect{y})$ between the prior and posterior entropy. We evaluate the IG for the oracle and the adaptive setup. Additionally, we report the IG for a \textit{static} setup where participants were assigned random items, in the same total amount as in the adaptive setup. In this static setup, participants thus answer the same number of questions on average (9.6), but items are not selected strategically.

We find $\text{IG}=0.79$ for the oracle (which sets the maximum amount of information that could be achieved), compared to $\mathrm{IG}=0.76$ for the adaptive setup. In other words, the adaptive setup retains $0.76/0.79=96\%$ of the information with 37\% fewer challenges. The trade-off between the information loss and the reduction in the number of trials can be adjusted by changing the early-stopping threshold $\epsilon$, with larger values associated with smaller information gains but also fewer trials. Finally, despite identical budget, the static design achieves a lower information gain. This means that the \gls{eig} maximization is effective and critical to the success of the strategy.

This shows that the active inference approach can be effective even in the absence of a prior calibrated item-bank. In fact, it has the potential to lower the costs of creating entirely new item banks on a set of subjects, by collecting calibration data adaptively. In some cases however, it might not be desirable to deploy this approach without having calibrated the items first; for instance, if the result of the testing procedure has any consequence for the subjects. Indeed, subjects tested at different points in the experiment are granted unequal opportunities, with late participants likely to enjoy more accurate results than early participants. Section \S\ref{section:leveraging} discusses how data from prior experiments, like calibration data for a bank of items, can be leveraged in this approach.  

\section{\label{section:adaptive_treatment}Experiment 2: adaptive treatment assignment}

\subsection{Introduction}

In this second experiment, we perform an adaptive search for optimal treatments (or conditions), i.e. treatments that maximize a desired outcome. In a conventional fixed design, we would typically divide the treatments evenly among participants. This can be inefficient, because this results in exploring the worst treatments and the best treatments just as often. In this context, active inference \citep{parr2022active} is an interesting approach that can incorporate pragmatic goals through a minor adjustment to \gls{eig}-driven optimization.  In this approach, treatments are allocated by weighing the associated \gls{eig} and expected utility, resulting in a careful balance between exploration and exploitation. We illustrate this technique step-by-step with an experiment seeking the treatments (in this case, the trivia questions) most associated with the education level of the participants. Such challenges could be interesting because they provide stereotypical examples of distinction, unveiling dimensions of cultural capital that can serve as reliable signals of social status \citep{bourdieu1984distinction,bourdieu1986forms}. In the context of psychology, this setup can support the search for discriminative stimuli that maximize differences in response between different groups, with potential applications in, say,  computerized adaptive diagnosis \citep{Gibbons2013,Gibbons2016} or cross-cultural research \citep{Barrett2020}. In this experiment, the budget is fixed (five treatments per participant). The active inference approach accurately recovers the optimal treatment up to approximately three times as often as a conventional design and competes with traditional algorithms.

\subsection{Methods}

The goal of this experiment is to find which questions best discriminate college-educated from non-college-educated participants, given a fixed budget of five treatments per participant (no stopping rule). Note that now, 
the same trivia questions from Experiment 1 now play the role of \textit{treatments}.  

The observation model  is a Beta--Bernoulli model with success probabilities $\phi_{dz}$ for each treatment $d$ and education group $z\in\{0,1\}$,
\begin{equation}
    y_{t+1,d}\mid z,\phi_{dz}\sim\operatorname{Bernoulli}(\phi_{dz}),
    \qquad
    \phi_{dz}\sim\operatorname{Beta}(1,1),
\end{equation}
where education $z=1$ for participants with a college education and $z=0$ for those without. 

The simulation model follows directly: for a candidate $d$ and the current participant's $z$, we draw $\phi_{dz}$ from its current Beta posterior and simulate a binary outcome.

In this experiment we also add preferences over outcomes. Let $y_{i,d}=1$ if participant $i$ answers item $d$ correctly, and $z_i=1$ if $i$ is college-educated. We want to encourage the sample to focus on exploring items that have promising levels of discrimination, rather than spending time on items suspected to be poorly discriminating. That preference is encoded as $\log p^{\ast}(y,z)=+\gamma$ if $y=z$ and $-\gamma$ otherwise, omitting the normalization constant for clarity. The utilitarian contribution $U(d)$ to the \gls{efe} then becomes:
\begin{equation}
    \label{eq:utility}
    \begin{aligned}
        U(d)\equiv \mathbb{E}_{q(y,z{\mid}d)}[\log p^{\ast}(y,z)]
        &= \gamma \left[q(y=z{\mid}d) - q(y\neq z{\mid}d)\right] + \text{cst} 
    \end{aligned}
\end{equation}.
Note that this pragmatic objective was missing from experiment 1.

Since we seek the design that maximizes $\mathrm{EIG}(d)+U(d)$,
$\gamma$ sets the exploration--exploitation tradeoff by weighing the contribution of the utility function in the objective (see Appendix~\S\ref{appendix:choosing_gamma}). When $\gamma \ll 1$, $p^{\ast}$ is uniform and active inference reduces to \gls{bad} again. When $\gamma\gg 1$, the expected utility dominates and active inference amounts to greedy utility maximization, which could quickly converge on poorly discriminative treatments. Therefore, a compromise must be found. Following \citet{Markovi2021}, we choose $\gamma=0.1$. Poor choices can undermine data acquisition, and Appendix \ref{appendix:choosing_gamma} discusses the rationale underlying the choice of $\gamma$ in detail.

Finally, because the model is conjugate, the posterior distribution for each parameter $\phi$ is exactly Beta distributed (with hyper-parameters reflecting success/failure counts) and the variational approximation can be skipped. We therefore provide exact derivations for these quantities in  Appendix~\ref{section:derivation_experiment_2}. More generally, users are free to avoid variational approximations if closed-form solutions exist.

\subsection{\label{section:treatment_data}Data}

We deployed this experiment on Prolific on the same 200 participants. For the simulation studies, we had to collect our own oracle on 200 more participants, since \citet{Dubourg2025} did not provide education data.

\subsection{\label{section:treatment_results}Results and evaluation}

\begin{figure}[!htp]
    \centering
    \begin{subfigure}[t]{0.45\textwidth}
        \centering
        \includegraphics[width=0.9\linewidth]{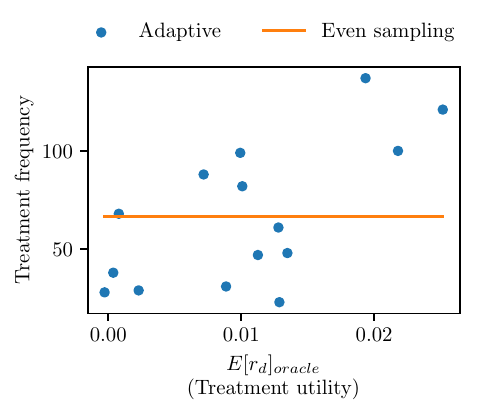}
        \caption{\label{fig:frequency}\textbf{Frequency of each treatment in the adaptive setup}. In the adaptive setup, superior treatments (according to the oracle) are administered more often on average. For comparison, the orange line shows the expected frequency of the treatments had they been divided evenly among participants, as in a conventional fixed design. }
    \end{subfigure}\hfill\begin{subfigure}[t]{0.45\textwidth}
        \centering
        \includegraphics[width=0.9\linewidth]{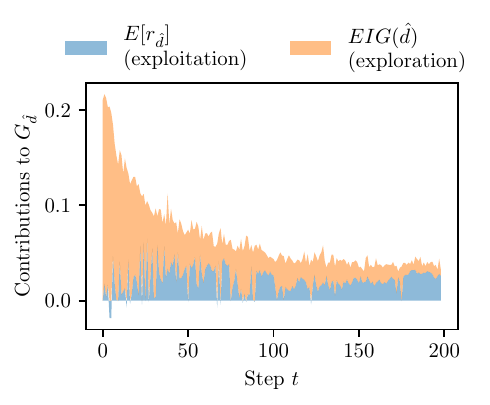}
        \caption{\label{fig:efe_composition}\textbf{Exploitation versus exploration during the experiment}.  The composition of the expected free energy of the chosen treatment is shown at each step of the experiment. Early in the experiment, the epistemic component (the EIG) dominates the \gls{efe}, and low-utility treatments are frequently selected, revealing the dominance of exploration. Late in the experiment, only high-utility treatments are selected, and exploitation prevails. }
    \end{subfigure}
     \begin{subfigure}{0.95\textwidth}
        \centering
        \includegraphics[width=1\linewidth]{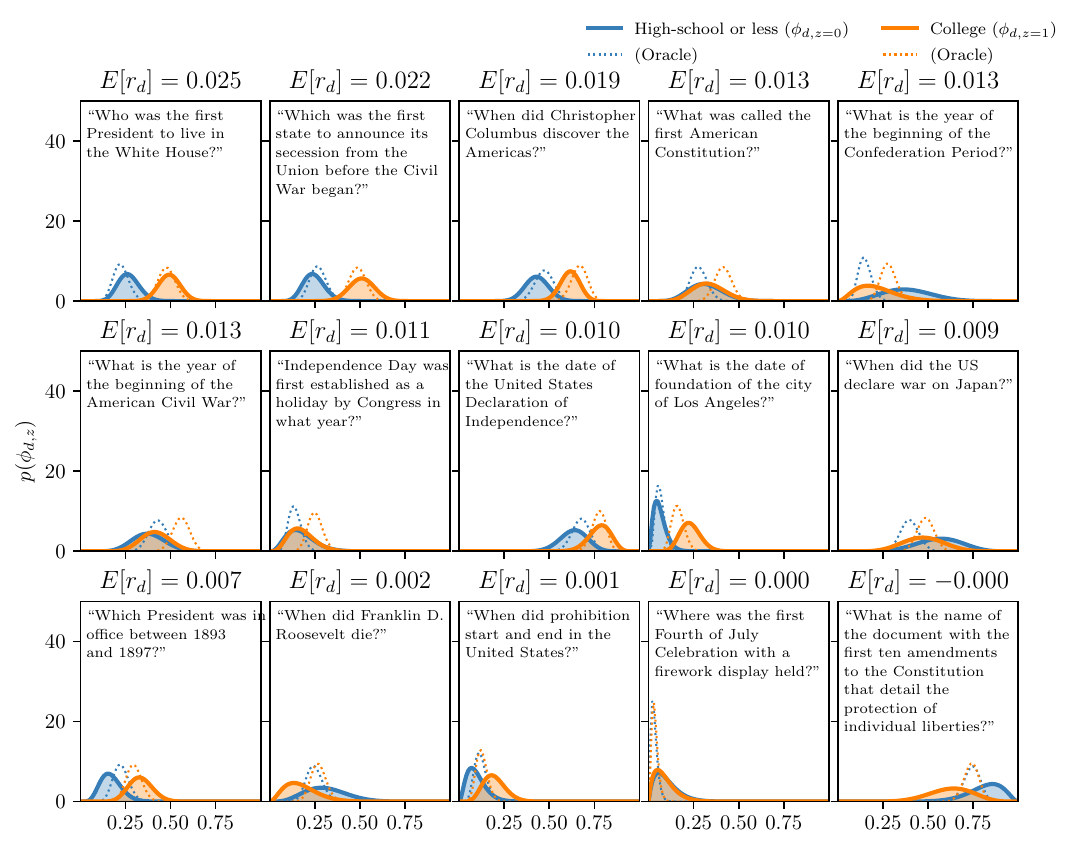}
        \caption{\label{fig:treatments}\textbf{Knowledge of each treatment (from best to worst) in the adaptive setup}.  }
    \end{subfigure}

    \caption{Simulation of the active inference approach.}
    \label{fig:active}
\end{figure}

We deployed and simulated the second experiment using 15 questions about American history. In the adaptive setup, we administered five ``treatments'' per participant using our active inference approach. Figure \ref{fig:frequency} shows that treatments with higher utility (per the oracle) tend to be delivered more often, with the best three treatments being delivered the most. Indeed, as shown in Figure \ref{fig:efe_composition}, as information about the treatment accumulates, the objective function that dictates which treatments are administered is increasingly influenced by the treatments' expected utility. This figure clearly depicts how active inference dynamically balances exploration and exploitation. Finally, Figure \ref{fig:treatments} compares the estimates of the success rate of participants with and without college education in the adaptive setup (solid lines) and in the oracle (dotted lines), from the best treatments (top-left) to the worst treatments (bottom-right). There is a close match between the oracle and the adaptive estimate for the best three treatments, despite 67\% fewer treatments being delivered to each participant in the adaptive setup. We find that the questions that best distinguish participants with different education levels are: ``Who was the first President to live in the White House?'', ``Which was the first state to announce its secession from the Union before the Civil War began?'', and ``When did Christopher Columbus discover the Americas?''.

To further assess the performance of the active inference setup, we measure the probability that it agrees with the oracle about which treatment is optimal. We include a comparison with a conventional fixed design (even sampling) and two reasonable strategies in adaptive treatment assignment. The first strategy is Thompson sampling, a highly popular bandit algorithm \citep{Russo2018}. This strategy amounts to drawing a treatment $d$ according to the probability $p_d$ that it is the superior treatment. In addition, we consider exploration sampling\footnote{In exploration sampling, treatment $d$ is selected with probability $p_d(1-p_d)/\sum_i(p_i(1-p_i))$, where $p_d$ is the Thompson sampling probability.}, a modification of Thompson sampling more suitable for adaptive experiments in policy-choice (\citealt{kasy2021adaptive})\footnote{More precisely, exploration sampling was designed to minimize the policy regret, i.e., the opportunity costs of enacting a suboptimal treatment after the conclusion of an experiment. }.

\begin{figure}
    \centering
    \includegraphics[width=0.9\linewidth]{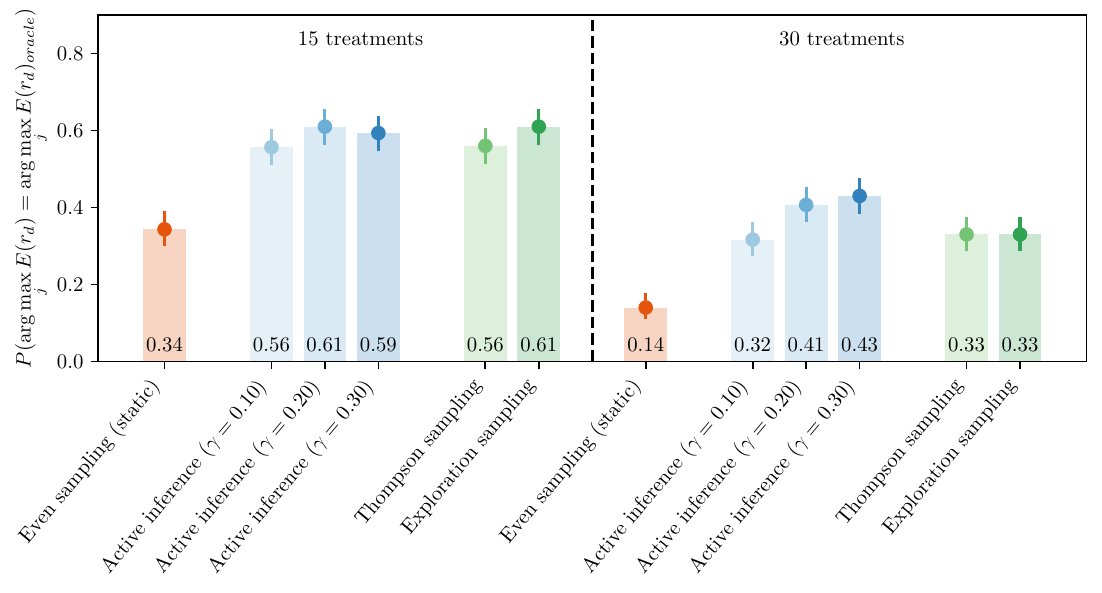}
    \caption{\textbf{Performance of active inference and traditional adaptive treatment assignment policies simulated on Experiment 2.} Performance is evaluated as the probability that the setup agrees with the oracle about which treatment is optimal (higher is better), given the expected utilities inferred from each setup and the expected utilities estimated from the oracle. The static design is included for reference. Each design is simulated 300 times on the oracle data. Error bars indicate 90\% credible intervals.}
    \label{fig:rl}
\end{figure}

We simulated each strategy 300 times using the oracle data, while randomizing the order of the participants in each run. For the active inference setup, we considered multiple preference priors, parameterized by different values of $\gamma$ (0.1, 0.2, and 0.3), where lower values privilege exploration, and larger values favor exploitation. We considered two scenarios: one including 15 treatments total (questions about American history), and one including 30 treatments total (questions about American history \& the solar system). The results are shown in Figure \ref{fig:rl}. These indicate that 1) adaptive strategies perform better; 2) performance is lower across the board when the treatment space is larger; 3) active inference can compete with and occasionally outperform Thompson and exploration sampling; 4) the performance of active inference depends on the choice of $\gamma$ (the prior preference over outcomes).

These specific results are based on a single underlying reward distribution (the oracle expected utilities) and should not be overgeneralized. However, prior evidence does indicate that active inference can outperform traditional bandit algorithms such as Thompson sampling when leveraging contextual information is advantageous \citep{Wakayama2025}. Active inference can indeed gather information more efficiently than many standard algorithms, owing to the \gls{eig} component of the expected free energy. In the present case, for instance, the information gained from administering a treatment depends on the education level $z$ of the participant; and in contrast to Thompson and exploration sampling, active inference does leverage this dependency in its decision-making process. In simpler settings, however, Thompson sampling is expected to perform better when evaluated against its own objective \citep{Markovi2021}. 

In order to further substantiate that these insights extend to adaptive experimentation more generally, we ran extensive simulations, varying the reward structure, the number of treatments, and the number of participants. The results confirm that active inference does well on several performance measures relevant to adaptive treatment assignment and can outperform specialized algorithms when the \gls{eig} term can guide the exploration more efficiently (Appendix \ref{appendix:active_inference_mab}). 

By controlling the tradeoff between exploration and exploitation, $\gamma$ affects how active inference ``performs'' according to a certain performance metric. Appendix \ref{appendix:choosing_gamma} proposes qualitative bounds on the parameter adjusting this tradeoff in the common case of Bernoulli multi-armed bandits. We must emphasize again that performance comparisons are difficult since active inference, Thompson sampling, and exploration sampling target different (and somewhat conflicting) performance criteria among which it might be difficult to choose \citep{horn2022comparison}. For example, the algorithms considered here are not specifically optimized for the performance metric used in Figure \ref{fig:rl} (known as best-arm identification in a bandit context; \citealt{russo2016simple}).

\section{\label{section:going-further}Going further}

Experiment 1 demonstrated the epistemic limit of active inference in an adaptive testing scenario. Experiment 2 considered a more general case where the utilitarian contribution was non-zero through an adaptive search for discriminative treatments. Despite different goals, both experiments use the same adaptive inference architecture and the same Python code. Additionally, in both cases, active inference achieves real-time capabilities and competed with or outperformed specialized traditional adaptive algorithms. However, since the examples covered in this paper are necessarily limited, we find it important to mention potential directions for addressing issues and needs that might arise in other applications. In particular, we discuss i) the early detection of model misspecification in real-time experiments (\S\ref{section:model_evaluation}); ii) the efficient exploitation of data from prior experiments (\S\ref{section:leveraging});  and iii) potential performance improvements for challenging problems (\S\ref{section:performance}).

\subsection{\label{section:model_evaluation}Real-time model evaluation}

Bayesian adaptive design relies on a statistical model of the data-generating process. If this model differs substantially from the actual process, the adaptive procedure may select inefficient or misguided designs, resulting in suboptimal data. Evaluating the model before data collection is often difficult; yet, in an adaptive experiment, data collection itself depends on the model. It is therefore important to monitor model adequacy as data are collected.

At a minimum, we must verify that the procedure correctly estimates the probability $p(y)$ of each possible outcome $y$ at the time the next challenge is selected. This quantity is particularly important here because both of our experiments rely on $p(y)$ when computing the \gls{efe}. Moreover, checking these predictions evaluates the complete predictive procedure, including both the statistical model and, when it is approximate, the inference method.

Concretely, for each delivered task, we record the outcome probabilities predicted by the model for this task before observing the participant's response. We then compare these predicted probabilities with the actual outcomes. For a binary outcome, for example, among trials for which the model predicts a probability of success close to $20\%$, participants should succeed approximately $20\%$ of the time. Because no predictions will be exactly equal to, say, $20\%$, we group predicted probabilities into eight equal-width bins, $[0,0.125), [0.125,0.25), \ldots$, and compare the mean predicted probability within each bin with the observed proportion of successes. This construction is called a \textit{reliability diagram} \citep{SilvaFilho2023}.

\begin{figure}[!htp]
    \centering
    \centering
    \begin{subfigure}[t]{0.475\textwidth}
    \centering
    \includegraphics[width=1\linewidth]{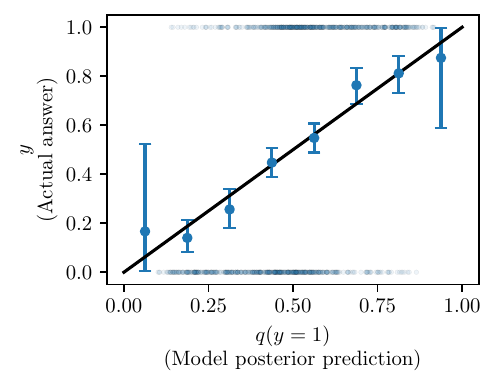}
    \caption{\textbf{Experiment 1}, using variational inference}
    \label{fig:evaluation_test}
    \end{subfigure}\hfill\begin{subfigure}[t]{0.475\textwidth}
    \centering
    \includegraphics[width=1\linewidth]{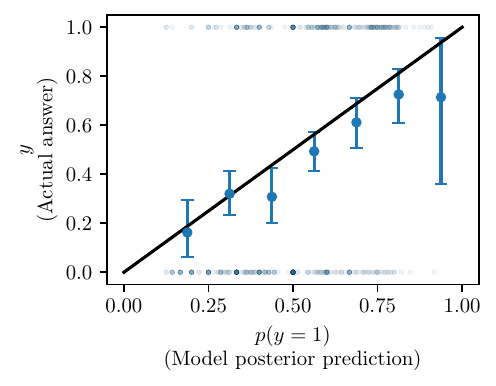}
    \caption{\textbf{Experiment 2}, using exact inference}
    \label{fig:evaluation_treatment}
    \end{subfigure}
    \caption{\textbf{Real-time model evaluation in Experiments 1 and 2, conducted halfway through the experiment}. The predicted probability $q(y=1)$ of the answer being correct (as estimated before administering the question) is contrasted to the actual answer $y\in\{0,1\}$. Error bars indicate the 95\% credible intervals of $\mathbb{E}[y]$ in each bin. Ideally, each bin should intersect with the black curve, which indicates perfect calibration.
        \label{fig:evaluation}}
\end{figure}

We show reliability diagrams for both our experiments in Figure \ref{fig:evaluation}, as evaluated halfway through each deployment. This reveals a good match between the predictions of the model and the observed outcomes. There is some indication that, in the first experiment (Figure \ref{fig:evaluation_test}), the model is slightly underconfident ($\bar{y}$ is closer to 0 or 1 than the model estimates). This suggests that a 2-parameter logistic item-response model, with sensitivity parameters $>1$, might be a better match to the data. Interestingly,  the quality of the predictions is slightly inferior in the second experiment (Figure \ref{fig:evaluation_treatment}). This mismatch may stem from inappropriate priors, or from the oversimplifying assumption that the probability of success of participants with a college degree is independent of the probability of success of those with no college degree. Therefore, even seemingly subtle modeling decisions should be subject to scrutiny when deploying adaptive experiments.

\subsection{\label{section:leveraging}Leveraging data from prior experiments}

In our case, the inference solely relies on data collected during the live experiment itself. In the context of computerized adaptive testing, this shows that the strategy is effective even when the tests have not been calibrated prior to deploying the experiment. However, in certain cases, we might want to leverage data from prior experiments (e.g. calibration data for a bank of items). This can be achieved in two ways. First, one can directly include data from prior experiments during the inference procedure. Second, if the volume of data makes it impractical and inefficient to learn from the entire data during the experiment, one can fit the model on previous experiments offline, recover the posterior distribution of each parameter (e.g., the items' difficulty), and use these posteriors as priors in the online experiment (see \S\ref{section:online_learning} on online learning).

\subsection{\label{section:performance}Performance and scalability improvements}

Experiments were implemented on AWS EC2 instances with 16 vCPU and 32 GB of memory (see Appendix \ref{appendix:performance}, Figure \ref{fig:performance} for further information about performance). We found that PsyNet could seamlessly support multiple participants in parallel, even in a demanding adaptive setup that requires propagating information between participants in real-time. Nevertheless, more difficult problems may require further optimization or alternative strategies. Below, we discuss ways in which higher computational efficiency may be achieved in these cases. 

\subsubsection{\label{section:online_learning}Online learning}

Reconstructing the posterior distribution from scratch at every step of the experiment (by learning from the entire data) can be inefficient in real-time computations. An appealing alternative is \textit{online learning}, which leverages the fact that all the relevant features of $\vect{y_{1:t}}$, the data accumulated at time $t$, are contained in the posterior  $p_t(\theta)=p(\theta{\mid}\vect{y_{1:t}})$. As a result, we can in principle obtain $p_{t+1}(\theta)$ simply by updating the posterior at time $t$ with the latest data $y_{t+1}$, which means computing $p_t(\theta{\mid}y_{t+1})$\footnote{\citet{vanderLinden2019} have leveraged a similar strategy to speed up tests in the context of Bayesian computerized adaptive testing.}. However, in the general case, the posterior distribution can only be reconstructed approximately (for instance, by relying on a variational approximation $q$). Relying on these approximations iteratively produces errors that can accumulate, resulting in substantial information loss or bias over time \citep{mussati2025prediction}. In some cases, however, the posterior distribution can be derived exactly and summarised by a few parameters that can be updated iteratively. This is the case for Experiment 2, in which all the relevant posterior distributions are Beta distributions that involve few parameters that can be updated iteratively upon the acquisition of new data. 

\subsubsection{Policy-based approaches and deep adaptive design}

Policy-based approaches to adaptive experiments use an appropriate mapping $\pi$ from the state/history of an experiment (e.g., $\vect{d_{1:t}}, \vect{y_{1:t}}$) to the optimal design to deliver at time $t+1$. The problem is then to find a policy that is fast to compute in real-time but nevertheless produces optimal decisions. Deep adaptive design \citep{foster2021deep} seeks policies $\pi_{\phi}$ that take the form of a neural network parameterized by the weights $\phi$. This has two interesting consequences. First, such policies are easy to evaluate in real time (no difficult Bayesian computation is necessary). Second, an optimal policy $\pi_{\hat{\phi}}$ can be approximated by simulating the experiments many times (by sampling from $p(\theta,y{\mid}d)$) and by performing stochastic gradient descent on a loss function capturing the total information gain achieved for a certain value of $\phi$. This policy-learning process can be performed offline, before running the actual experiment, when the computational cost is not a concern. This line of approach is thus probably most suitable in difficult cases, where variational inference does not produce sufficient speed-ups. Users interested in this technique may also benefit from using PsyNet, because this framework makes it easier to simulate and deploy models with the same code.

In addition, deep adaptive design can outperform traditional \gls{bad} because it is naturally non-greedy: rather than maximizing the \gls{eig} step by step, it can deliver sequences of tasks that improve \gls{eig} in the long run \citep{rainforth2023modernbayesianexperimentaldesign}. However, potential limitations may arise in discrete design spaces, due to the reliance on gradient-based methods \citep{foster2019variational}, although related developments include solutions to adaptive testing with discrete item-banks \citep{Li2026}. Synergies between deep policy learning and active inference have been explored \citep{Millidge2020}, which suggests potential connections with deep adaptive design.

\section{Conclusion}

This paper provided a concrete introduction to a flexible and efficient approach to real-time adaptive experiments. Literature on adaptive experimentation remains fragmented across disciplines, with areas such as computerized adaptive testing and adaptive treatment assignment developing largely independent methodological traditions. This fragmentation prevents researchers from leveraging advances in adjacent fields, leading to duplicated efforts and missed opportunities. In addition, adaptive experiments entail computational and technical obstacles that can make their implementation challenging. We address these issues by introducing a practical and general approach to real-time adaptive experiments that bridges different traditions. Our solution combines three key components: active inference, a Bayesian computational framework inspired by cognitive neuroscience; Pyro, a probabilistic programming library; and PsyNet, a software package for large-scale online experiments.

While active inference was developed as a theory of perception and action in biological systems, it interestingly provides a minimal mathematical toolbox that can cover a wide range of scenarios in adaptive experimentation. While traditional approaches to adaptive experimentation either optimize purely epistemic objectives (by maximizing the information gain) or purely pragmatic ones (maximizing an expected utility), active inference provides a principled framework that inherently combines both motivations. Consequently, active inference covers a broad range of situations, from computerized adaptive testing to adaptive treatment assignment. In the context of real-time experiments, in which the optimal design must be computed efficiently, active inference provides several advantages. First, it naturally entails variational inference schemes for achieving faster (albeit approximate) inference. In addition, by incorporating the expected information gain in its decision-making, active inference generates reasonably efficient exploration strategies without necessarily explicitly factoring future steps in its calculations (although doing so may bring improvements and remains a theoretical possibility; \citealt{Paul2024}). Active inference proves especially valuable whenever high modeling flexibility is important, owing to its broadly applicable objective function that can accommodate epistemic and pragmatic considerations. However, for specific cases where well-calibrated standard solutions exist (including computerized adaptive testing with simple item-response models), the advantage is less marked, although we see clear benefits in adaptive testing without calibrated items. In such scenarios, active inference (and traditional \gls{bad}) can improve learning by managing large prior uncertainty on the items' difficulty efficiently (see Appendix \ref{appendix:eig_2pl}, Figure \ref{fig:eig_vs_fisher}), and even progressively learn to dismiss tests with undesirable features like slow response times. 

The flexibility of active inference is best leveraged in conjunction with probabilistic programming libraries. For instance, Pyro makes the approach easily deployable in new contexts with limited effort and model-specific adjustments, allowing researchers to focus on their own assumptions rather than implementation details. Specifically, we recommend Pyro because it natively implements variational inference schemes and several \gls{eig} estimation algorithms. However, users can easily replace it with a library of their choice in our approach.

Finally, to fully leverage the potential of this framework in real-time applications, we suggest the use of PsyNet, a platform for large-scale online experiments that greatly facilitates the adoption of state-of-the-art techniques in Bayesian adaptive design. As a Python framework, PsyNet can leverage the most recent probabilistic programming libraries that make Bayesian approaches scalable and sufficiently fast for real-time experiments. In addition, just like Bayesian approaches, PsyNet is agnostic about the precise nature of the task; it supports stimuli and responses in arbitrary domains, such as speech, music, video, and more, including in the context of complex social experiments. Interestingly, the modular structure of PsyNet (as an object-oriented framework) makes it easy to make drastic changes to an experiment (e.g., by changing the domain of the task, or, in adaptive experiments, the optimization strategy) with minimal code changes. Such modularity is also conducive to promoting exchanges of methods and code across the behavioral sciences. PsyNet provides a practical advantage for large-scale online adaptive experiments, in part because of its ability to process several participants in parallel, thus considerably speeding up data collection for experiments involving thousands of subjects. PsyNet is also attractive due to its ability to support a broad range of stimulus and response types, including audio, video, multiple choice, free text, sliders, and WebGL. Finally, PsyNet can easily accommodate adaptive strategies other than active inference, like for instance traditional computerized adaptive testing, especially extant Python implementations \citep{Engicht2026}. The comparative advantage of PsyNet increases with the complexity of the experimental design, the need for modularity, and/or the need to rely on Python libraries that JavaScript framework cannot integrate easily.

While technical barriers remain, recent advances in agentic AI may lower implementation costs. The modular structure of the approach described here appears particularly amenable to such automation: probabilistic programming provides a convenient interface for large language models to translate statistical assumptions into executable models, while frameworks such as PsyNet may facilitate the implementation of experimental paradigms. This suggests the potential relevance of this framework for the broader development of automated, closed-loop scientific discovery \citep{Musslick2024,auto-psych,ashkay}, in which computational systems may eventually assist not only with optimizing data collection within experiments, but also with designing and selecting experiments themselves.

\printglossary[type=\acronymtype]

\section*{Statements}

\paragraph*{Acknowledgments}

We thank the members of the NSF project \textit{Designing smart environments to augment collective learning \& creativity} for their feedback. We also thank Wesley Bonifay, Clintin Stober, and Sonja Winter for useful discussions.

\paragraph*{Funding}

This work was supported by the NSF grant ``Collaborative Research: Designing smart environments to augment collective learning \& creativity'' (award 171418).

\paragraph*{Conflicts of interests}

The authors declare no competing interests.

\paragraph*{Ethics approval}

Recruitment was done in compliance with approved Max Planck Society Ethics Council protocols (2021-42).

\paragraph*{Consent to participate}

Informed consent was obtained from all individual participants included in the study.

\paragraph*{Consent for publication}

Not applicable.

\paragraph*{Availability of data and material}

All the data are available at \codeurl.

\paragraph*{Code availability}

The code is available at \codeurl.

\printbibliography

\appendix 

\section{\label{appendix:workflow_details}Technical details for the active inference workflow}

\subsection{\label{section:efe}Expected free energy}

Following active inference, we deliver the experimental design that minimizes the expected free energy $G(d)$. This quantity can be expressed as follows:

\begin{alignat}{3}
    G(d) &= &&-\mathbb{E}_{q(y,\theta{\mid}d)}\left[\log\dfrac{q(y{\mid}\theta,d)}{q(y{\mid}d)}\right] &&- \mathbb{E}_{q(y{\mid}d)} \left[ \log p^{\ast}(y)\right]\label{eq:efe}\\
    \underset{q\to p}&{\simeq} &&-\underbrace{\mathrm{EIG}(d)}_{\substack{\text{information-seeking}\\\text{intrinsic/epistemic motivation}\\\text{\textit{(exploration)}}}} &&-\underbrace{U(d)}_{\substack{\text{utilitarian value}\\\text{extrinsic/pragmatic motivation}\\\text{\textit{(exploitation)}}}}\label{eq:efe_simple}
\end{alignat}

Let us unpack the two terms in the right-hand side of \eqref{eq:efe}. In both of these, $q$ denotes approximations of the relevant probability distributions $p$.

The first is equal to the EIG when $q=p$. Indeed, setting $q = p$ in the first term of \eqref{eq:efe}, and writing
$p(y,\theta{\mid}d) = p(y{\mid}\theta,d)\,p(\theta)$:
\begin{align}
\mathbb{E}_{p(y,\theta{\mid}d)}\!\left[\log \frac{p(y{\mid}\theta,d)}{p(y{\mid}d)}\right]
&= \mathbb{E}_{p(y,\theta{\mid}d)}\!\left[\log \frac{p(y,\theta{\mid}d)}{p(y{\mid}d)\,p(\theta)}\right]
  \nonumber\\
&= \mathbb{E}_{p(y,\theta{\mid}d)}\!\left[\log \frac{p(\theta{\mid}y,d)}{p(\theta)}\right]
  \nonumber\\
&= \mathbb{E}_{p(y{\mid}d)}\Big[\,\mathbb{E}_{p(\theta{\mid}y,d)}\big[\log p(\theta{\mid}y,d)\big]\Big]
   - \mathbb{E}_{p(\theta)}\big[\log p(\theta)\big] \nonumber\\
&= -\,H(\theta \mid y, d) + H(\theta) \nonumber\\
&= \mathrm{EIG}(d),
\end{align}
where the second-to-last line uses the definition of the (conditional)
entropy, $H(\theta) = -\mathbb{E}_{p(\theta)}[\log p(\theta)]$ and
$H(\theta{\mid}y,d) = -\mathbb{E}_{p(y{\mid}d)}\mathbb{E}_{p(\theta{\mid}y,d)}[\log p(\theta{\mid}y,d)]$.

In the second term of \eqref{eq:efe}, $p^{\ast}(y)$ is a prior distribution on $y$ that measures our preference for/aversion to certain outcomes (for instance, if we prefer $y=1$ over $y=0$, we would have $p^{\ast}(y=1)>p^{\ast}(y=0)$). We can also regard the second term as the expected utility of design $d$, given a reward function $r(y)=\log p^{\ast}(y)$.

To construct the approximate distributions $q$ that appear in \eqref{eq:efe}, active inference bypasses difficult computations by appealing to variational inference.


\subsection{\label{section:variational}Variational inference for posterior estimation}

The state of knowledge in the experiment at a step $t$ is reflected by the posterior distribution $p(\vect{\theta}{\mid}\vect{y}_{1:t})$, where $\theta$ can reflect, for instance, the participants' ability and the challenges' difficulty level given all prior answers $1\dots t$. Making informed decisions requires learning this posterior distribution iteratively in real-time. In general however, this distribution admits no closed-form expression. Thus, we must approximate the true posterior in another way. In the framework of active inference, agents continuously update their beliefs about the world via variational inference. While it provides less guarantee compared to traditional MCMC methods, this technique can speed up inference, especially for larger amounts of data \citep{Blei2017,advi,svi}. In a nutshell, variational inference seeks an approximation $q_{\vect{\phi}}(\vect{\theta})$ to the posterior $p(\vect{\theta}{\mid}\vect{y}_{1:t})$ by drawing from a family of distributions continuously parameterized by $\phi$. For instance, a common strategy is to approximate $p(\vect{\theta}{\mid}\vect{y}_{1:t})$ by independent normal distributions, such that $q(\vect{\theta})=\prod_i  q_i(\theta_i)$. Then, $q_{\vect{\phi}}$ can be parameterized by the means $\vect{\mu}$ and standard deviations $\vect{\sigma}$ of these distributions -- i.e., $\vect{\phi}=(\vect{\mu},\vect{\sigma})$. The best approximation is achieved by minimizing the divergence between $q_{\vect{\phi}}(\vect{\theta})$ and $p(\vect{\theta}{\mid}\vect{y}_{1:t})$:

\begin{align}
    \hat{\phi} &\equiv \argmin_{\vect{\phi}} \underbrace{D_{KL}(q_{\vect{\phi}}(\vect{\theta}){\mid}{\mid}p(\vect{\theta}{\mid}\vect{y}_{1:t}))}_{\substack{\text{Divergence between approximation}\\\text{and true posterior}\\\text{(intractable)}}}\\ &= \argmin_{\vect{\phi}} \mathbb{E}_{q_{\phi}(\vect{\theta})}\left[\log \dfrac{q_{\phi}(\vect{\theta})}{p(\vect{\theta}{\mid}\vect{y}_{1:t})}\right] \text{(by definition of } D_{KL})
\end{align}

\noindent where $D_{KL}(q,p)=0$ if $p=q$. 
Unfortunately, the divergence is intractable, since $p(\vect{\theta}{\mid}\vect{y}_{1:t})$ is itself intractable. However, we can decompose it into two terms:

\begin{align}
    \hat{\phi} &= \argmin_{\vect{\phi}} \mathbb{E}_{q_{\phi}(\vect{\theta})}\left[\log \dfrac{q_{\phi}(\vect{\theta})}{p(\vect{y}_{1:t},\vect{\theta})}  +  \log p(\vect{y}_{1:t}) \right] 
\end{align}

\noindent where $p(y_{1:t}, \theta)=p(y_{1:t}{\mid} \theta)p(\theta)$ is often, as in the present case, tractable. The first term, known as the variational free energy, is therefore computationally manageable. We can drop the second term, since it is constant. We can then obtain $\hat{q}$ by minimizing the variational free energy:

\begin{equation}
    \label{eq:vfe}
    \hat{\phi} = \argmin_{\vect{\phi}} \underbrace{ \mathbb{E}_{q_{\phi}(\vect{\theta})}\left[\log \dfrac{q_{\phi}(\vect{\theta})}{p(\vect{y}_{1:t},\vect{\theta})}\right] }_{\substack{\text{Variational free energy}}}
\end{equation}

Thus, learning the posterior becomes an optimization problem \citep{Blei2017} which can be solved efficiently by methods such as stochastic gradient descent \citep{wingate2013automated}. In active inference, computations leverage the variational approximations of intractable posterior distributions achieved via the minimization of the variational free energy, replacing $p_t(\theta)=p(\theta{\mid}y_{1:t},d_{1:t})$ with $\hat{q}(\theta)$. For instance, we may assume that $p_t(y,\theta{\mid}d)\equiv p_t(y{\mid}\theta,d)p(\theta{\mid}y_{1:t},d_{1:t}) \simeq p(y{\mid}\theta,d)q(\theta)$, where $q$ is the approximation of the intractable posterior of $\theta$ at time $t$. 

\subsection{\label{section:variational_eig}Estimating the expected information gain (EIG)}

In active inference (as in more traditional \gls{bad}), the expected information gain is a key quantity in determining an optimal design. The expected information gain is simply the difference between the entropy $H(\theta)$ (the uncertainty about $\theta$ before observing a participant's response) and $H(\theta{\mid}y_{i,d})$, the expected uncertainty after having observed it (also known as the conditional entropy). It can be expressed in the following ways\footnote{\eqref{eq:eig} $\to$ \eqref{eq:eig_post} follows from the definition of the entropy, $H(x)=-\mathbb{E}_{p(x)}[\log p(x)]$, and \eqref{eq:eig_post}$\to$\eqref{eq:eig_marginal} follows from the Bayes rule, $p(\theta{\mid}y)=\dfrac{p(y{\mid}\theta)p(\theta)}{p(y)}$. The latter form is convenient since it avoids computing the posterior $p(\theta{\mid}y_{i,d})$ for each possible outcome \citep{foster2021a}.}:

\begin{align}
    \label{eq:eig}
    \text{EIG}(d) &\equiv H(\theta)-H(\theta{\mid}y_{t+1},d)\\
    &= \mathbb{E}_{p_t(y_{t+1},\theta{\mid}d)} \left( \log \dfrac{p_t(\theta{\mid}y_{t+1},d) }{p_t(\theta{\mid}d)} \right)\label{eq:eig_post}\\
    &= \mathbb{E}_{p_t(y_{t+1},\theta{\mid}d)} \left( \log \dfrac{p_t(y_{t+1}{\mid}\theta,d) }{p_t(y_{t+1}{\mid}d)} \right)\label{eq:eig_marginal}
\end{align}

\noindent where the subscript $t$ in $p_t$ designates probability distributions conditional on all prior data, i.e. $p_t(\cdot)\equiv p(\cdot{\mid}y_{1:t},d_{1:t})$.

The optimal challenge is the one with the highest EIG; we must thus estimate the EIG for each challenge $d$ and choose the maximum:

\begin{equation}
    \hat{d} = \argmax_d \text{EIG}(d)
\end{equation}

Even for a single design however, evaluating the EIG can be difficult. A naive approach would be to estimate \eqref{eq:eig_marginal} by drawing values of $(y,\theta)$ from $p(y{\mid}\theta,d)q(\theta)$, the variational approximation of $p_t(y,\theta)$ and calculating the mean of the integrand:

\begin{equation}
    \label{eq:eig_estimate}
    \mathrm{EIG}(d) \simeq \dfrac{1}{S} \sum_{s=1}^S \log \dfrac{p_t(y_s{\mid}\theta_s,d)}{p_t(y_s{\mid}d)}
\end{equation}

Unfortunately, in most cases, the marginal probability $p_t(y_s{\mid}d)$ is itself intractable and requires its own approximation. The conventional approach in those cases is to employ a nested estimator\footnote{Specifically:

\begin{equation}
    \mathrm{EIG}(d) \simeq \dfrac{1}{N} \sum_{s=1}^N \log \dfrac{p_t(y_s{\mid}\theta_s,d)}{\dfrac{1}{M} \sum_{\tilde{s}=1}^M p(y_{s}{\mid}\tilde{\theta}_{\tilde{s}},d)}
\end{equation}

However, estimating this quantity would require $\mathcal{O}(M\times N)$ computations, which can be challenging in real-time experiments. The computational cost depends on the particular values of $M$ and $N$. Under broad assumptions, this estimator converges as $\mathcal{O}(\sqrt{1/N + 1/M^2})$, and it is optimal for $M$ to scale as $\sqrt{N}$, requiring $\mathcal{O}(N^{3/2})$ computations in practice. We refer readers interested in this technique to prior literature \citep{rainforth2018nesting}.}. %

Instead, below, for the sake of generality and consistency with active inference, we appeal to a faster variational approximation scheme from \citet{foster2019variational}. This approach minimizes an upper-bound to the \gls{eig} rather than the \gls{eig} itself, noticing that:
\begin{equation}
    \label{eq:eig_variational}
    \text{EIG}(d)\leq \underbrace{\mathbb{E}_{p_t(y_{t+1},\theta{\mid}d)}\left(\log\dfrac{ p(y_{t+1}{\mid}\theta,d)}{q_{\hat{\phi}}(y_{t+1}{\mid}d)}\right)}_{\text{upper-bound}} \simeq \dfrac{1}{S} \sum_{s=1}^S \log \dfrac{p(y_{s}{\mid}\theta_{s},d)}{q(y_{s}{\mid}d)}
\end{equation}

\noindent where $q(y{\mid}d)$ is a variational approximation of $p_t(y_{t+1}{\mid}d)$, the marginal distribution of $y$, and the samples are drawn from $p_t(y,\theta{\mid}d)$. The upper bound in \eqref{eq:eig_variational} is equal to the EIG when $q=p$. The best approximation $q_{\hat{\phi}}$ is obtained by minimizing the upper-bound, using the same strategy as previously employed in the approximation of the posterior distribution $p_t(\theta)$. 

This approach provides several advantages. First, it provides more computationally favorable convergence rates than the conventional nested estimator ($\mathcal{O}(T^{-1/2})$ instead of $\mathcal{O}(T^{-1/3})$, for $T$ computations). Second, it offers both conceptual and practical alignment with active inference. In particular, this technique jointly provides two quantities crucial to active inference in one go: the \gls{eig}, \textit{and} a variational approximation $q(y)$ of the marginal distribution of $y$ -- which enters the pragmatic term in the \gls{efe}, when pragmatic considerations apply.

In the case of Experiment 1, we choose $q_{\phi_d}(y{\mid}d) = \operatorname{Bernoulli}(y,\text{logit}^{-1}(\phi_d))$, which is well-suited for a binary variable such as $y$. We use a Pyro function (\python{marginal_eig()}) that directly implements this approach \citep{foster2019variational,pyro_working_memory_tutorial}. The implementation requires choices about the number of optimization steps and the number of samples drawn from the variational approximation. In Experiment 1, we used 400 optimization steps with 400 samples per step, followed by $S=10\,000$ samples for the final Monte Carlo estimator. These values were selected to provide a satisfactory compromise between accuracy and computational performance. For other applications, we recommend benchmarking these parameters on simulated data. If accuracy is insufficient, the relevant parameters can be increased separately to identify the bottleneck; if computation is too slow, we recommend first reducing the number of optimization steps, potentially together with a higher learning rate.

Finally, let us remark that EIG maximization shares some resemblance with maximal Fisher information criterion, which is widely used in computerized adaptive testing. However, these are not equivalent \citep{rainforth2023modernbayesianexperimentaldesign}. Appendix \ref{appendix:eig_2pl} compares the two criteria in another item-response model common in adaptive testing literature (2PL).  

\section{\label{section:derivation_experiment_2}Exact derivations of the posterior and the EIG in Experiment 2}

While crucial to our computations, $q(y_{i,d},z_i{\mid}d)$ is not known a priori: it is only learned through the accumulation of observations ($\vect{y_{1:t}},\vect{z}_{1:t}$) resulting from ``treating'' the participants with the trivia questions. Our task is thus to estimate $q$ given prior observations. First of all, let us factorize it as $q(y_{i,d},z_i{\mid}d)=q(y_{i,d}{\mid}z_{i},d)q(z_i)$. Thus we decompose the problem of computing the posterior predictive distribution into the independent computation of $q(y_{i,d}{\mid}z_{i},d)$ (the probability that a participant answers correctly or not to $d$, given their education) and $q(z_{i})$ (the marginal frequency of each education level in the pool of participants). The latter is easy, since we recruited our participants such that $q(z_{i})=1/2$ (half of them are college-educated).

There remains to infer $q(y_{i,d}{\mid}z_{i},d)$. In our case, we can compute this distribution without any variational approximation $q$ and all the posteriors below will be exact\footnote{In general, the posterior distribution is difficult to compute, and we would often need to appeal to approximate methods such as variational inference, as in \ref{section:variational}.}. We assume that $y_{i,d}{\mid}z_i,\phi_{d,z_i}\sim\mathrm{Bernoulli}(\phi_{d,z_i})$, where $\phi_{dz} \in [0,1]$ is the unknown probability that a participant with education level $z$ answers correctly to $d$. The number of successes ($y=1$) follows a binomial distribution with success probability $\phi_{d,z_i}$. For a given treatment $d$ and education group $z$, let
\[
    \mathcal{I}_{d,z}
    =
    \left\{
        i : z_i=z
        \text{ and participant } i \text{ received treatment } d
    \right\}
\]
denote the set of participants in group $z$ who were administered treatment $d$.
The number of correct responses to treatment $d$ in group $z$ therefore follows
\begin{equation}
    \sum_{i\in\mathcal{I}_{d,z}} y_{i,d}
    \mid
    \phi_{d,z}
    \sim
    \operatorname{Binomial}
    \left(
        {\mid}\mathcal{I}_{d,z}{\mid},
        \phi_{d,z}
    \right).
\end{equation}

Assuming the uniform prior
$\phi_{d,z}\sim\operatorname{Beta}(1,1)$,
conjugacy gives the posterior
\begin{equation}
    \phi_{d,z}
    \mid \vect{y}_{1:t},\vect{z}_{1:t}
    \sim
    \operatorname{Beta}
    \left(
        \alpha_{d,z},
        \beta_{d,z}
    \right),
\end{equation}
where
\begin{align}
    \alpha_{d,z}
    &= 1 + \sum_{i\in\mathcal{I}_{d,z}} y_{i,d}, \\
    \beta_{d,z}
    &= 1 + \sum_{i\in\mathcal{I}_{d,z}} (1-y_{i,d}).
\end{align}

The posterior predictive probability of a correct response is therefore:

\begin{equation}
     p(y_{i,d}=1 \mid z_i=z,d)
    =
    \mathbb{E}_{p(\phi_{d,z})}
    \left[
        p(y_{i,d}=1 \mid z_i=z,d,\phi_{d,z})
    \right]
    =
    \frac{\alpha_{d,z}}
         {\alpha_{d,z}+\beta_{d,z}}.
\end{equation}

We are now ready to calculate \eqref{eq:utility}, the ``utilitarian'' contribution to the \gls{efe}, as follows:

\begin{equation}
\begin{aligned}
\mathbb{E}_{q(y,z\mid d)}[\log p^{\ast}(y,z)]
&=
\gamma \left[
\overbrace{
    \frac{1}{2}p(y=1\mid z=1,d)
    +\frac{1}{2}p(y=0\mid z=0,d)
}^{p(y=z\mid d)}
\right. \\
&\qquad\left.
-
\underbrace{
    \frac{1}{2}p(y=0\mid z=1,d)
    +\frac{1}{2}p(y=1\mid z=0,d)
}_{p(y\neq z\mid d)}
\right]
+\mathrm{cst}
\\
&=
\gamma\left[
\frac{\alpha_{d,1}}{\alpha_{d,1}+\beta_{d,1}}
-
\frac{\alpha_{d,0}}{\alpha_{d,0}+\beta_{d,0}}
\right]
+\mathrm{cst}.
\end{aligned}
\end{equation}

Finally, in order to evaluate the objective \eqref{eq:efe}, we still need to compute the expected information gain, which in our case amounts to:

\begin{equation}
    \mathrm{EIG}(d) = \mathbb{E}_{p(y_{i,d},\phi_{d,z_i}{\mid}d,z_i)} \left[\log\dfrac{p(y_{i,d}{\mid}\phi_{d,z_i},d,z_i)}{p(y_{i,d}{\mid}d,z_i)}\right]
\end{equation}

It would be natural to approximate this quantity through Monte Carlo, by drawing $S$ samples of $y_{i,d}$ and $\phi_{d,z_i}$ from  $p(y_{i,d},\phi_{d,z_i}{\mid}d,z_i)$:

\begin{equation}
    \mathrm{EIG}(d) \simeq \dfrac{1}{S} \sum_{s=1}^S \left[\log\dfrac{y_{i,ds}\phi_{d,z_i,s} + (1-y_{i,ds})(1-\phi_{d,z_i,s})}{y_{i,ds}\dfrac{\alpha_{d,z_i}}{\alpha_{d,z_i}+\beta_{d,z_i}} + (1-y_{i,ds})\dfrac{\beta_{d,z_i}}{\alpha_{d,z_i}+\beta_{d,z_i}}}\right]
\end{equation}

In the rather common case of Beta-Bernoulli multi-armed bandits (to which our example is closely related), the \gls{eig} in fact admits an analytical expression, specifically (omitting the $z_i$ indices for readability):

\begin{equation}
    \label{eq:analytical_eig}
    \mathrm{EIG}(d)=-\mu_d\log\mu_d-(1-\mu_d)\log(1-\mu_d)+\mu_d\psi(\mu_dn_d+1)+(1-\mu_d)\psi((1-\mu_d)n_d+1)-\psi(n_d+1)
\end{equation}

\noindent where $\mu_d\equiv\alpha_d/(\alpha_d+\beta_d)$, $n_d=\alpha_d+\beta_d$, and $\psi$ is the digamma function.  To maintain a higher level of generality, our tutorial implements the Monte-Carlo estimator instead of this analytical expression (with $S=5000$); however, \eqref{eq:analytical_eig} can be relevant to a broad class of experiments featuring treatments with binary outcomes.

\section{\label{appendix:eig_2pl}Comparing EIG maximization and Fisher information maximization in computerized adaptive testing}

\begin{figure}[H]
    \centering
    \includegraphics[width=0.9\linewidth]{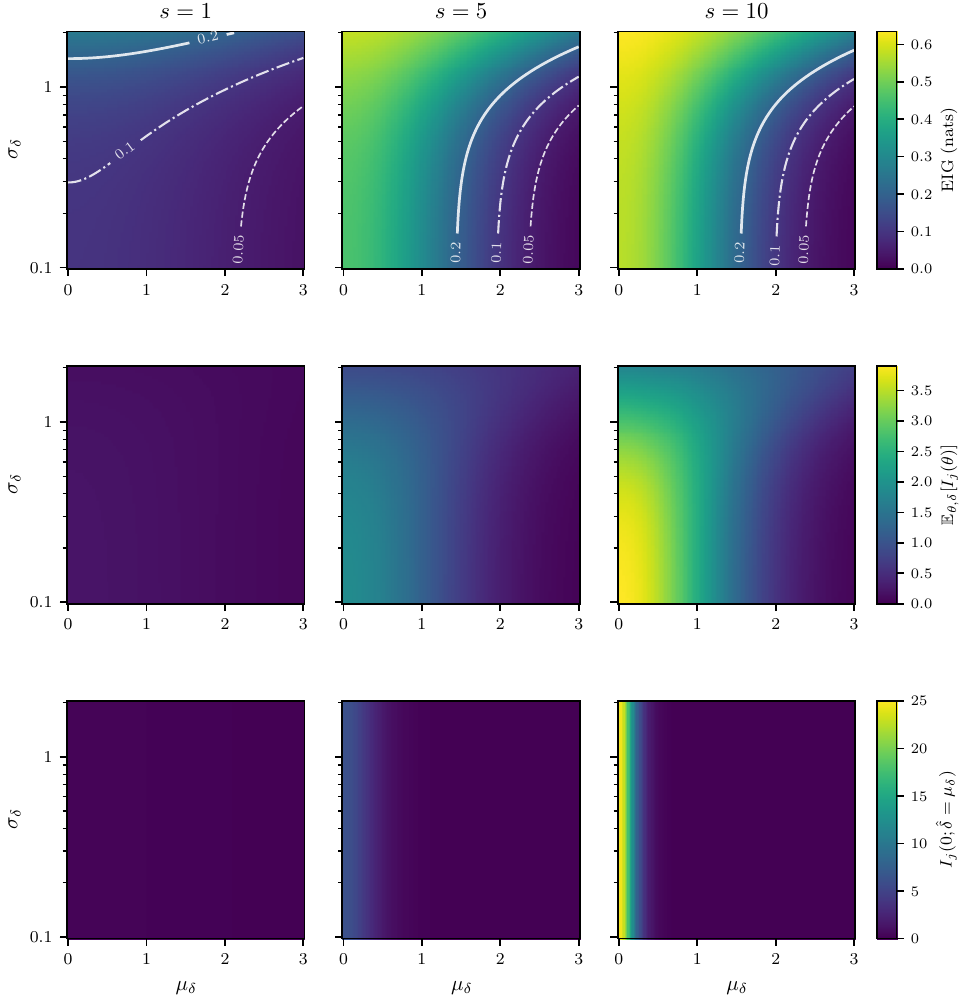}
    \caption{\textbf{Expected information gain (upper plots) versus the expected Fisher information (intermediate plots) and point-wise Fisher information (lower plots) as a function of the test parameters in the 2PL model}. In this model, $p(y=1{\mid}\theta,\delta,s)=\mathrm{logit}^{-1}(s(\theta-\delta))$, with $\theta$ the participant ability, $\delta$ the test difficulty, and $s$ the test sensitivity. We assume $\theta\sim\mathcal{N}(0,1)$ and $\delta\sim\mathcal{N}(\mu_{\delta},\sigma_{\delta})$. }
    \label{fig:eig_2pl}
\end{figure}

\begin{figure}[H]
    \centering
    \includegraphics[width=0.67\linewidth]{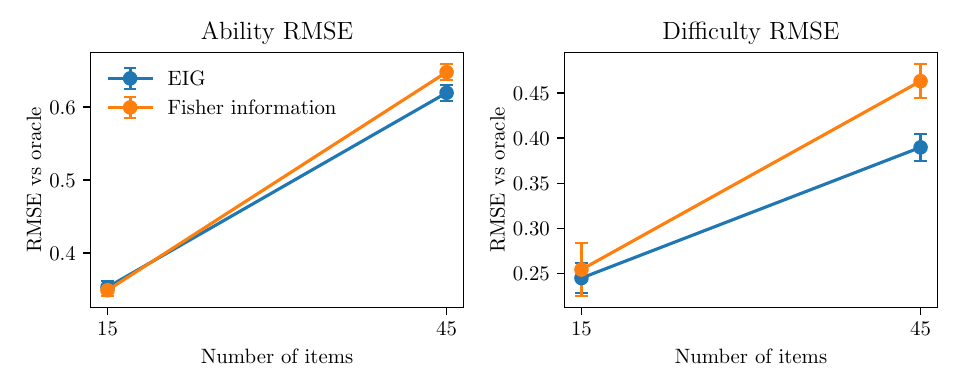}
    \caption{Root mean squared error (RMSE) of ability $(\theta)$ and difficulty ($\delta$) estimates in simulations of computerized adaptive testing under EIG and Fisher information maximization, for 15 and 45 items, 200 participants, a fixed stopping rule (10 tasks per participant), using 8 simulations for each strategy with paired random abilities ($\mathcal{N}(0,1)$) and difficulties ($\mathcal{U}(-2,+2)$). For large numbers of items, EIG-maximization significantly improves the accuracy of difficulty estimates.}
    \label{fig:eig_vs_fisher}
\end{figure}

\section{\label{appendix:slow} Eliminating slow tests with active inference}

By combining the \gls{eig} and the expected utility, the expected free energy embeds a cost-benefit analysis contrasting the amount of information gained from an action and the cost of performing this action. Take, for instance, a computerized adaptive testing scenario. This sort of experiment is purely driven by the maximization of knowledge about the participants via the \gls{eig}. But suppose we would like to weigh the \gls{eig} against the pragmatic cost of tests that take too long to complete. In active inference, the pragmatic cost of an outcome $x$ is implemented in the \gls{efe} by a contribution $\mathbb{E}_{q(x{\mid}d)}[\log p^{\ast}(x)]$, where $p^{\ast}(x)$ represents prior preferences over $x$. To incorporate the cost of slow tests in our decision making, we must find an appropriate prior preference $p^{\ast}(\tau_d)$ where $\tau_d$ is the time it takes to complete test $d$. An interesting candidate is the following improper prior:

\begin{equation}
    p^{\ast}(\tau) = \begin{cases}
      1 & \text{ if } \tau \leq \tau_{\mathrm{slow}}\\
      e^{-\kappa} & \text{otherwise}
    \end{cases}
\end{equation}

Under such a prior, the expected utility term in the \gls{efe} reduces to:

\begin{equation}
    \mathbb{E}_{q(\tau_d)}[\log p^{\ast}(\tau_d)] = -\kappa \int_{\tau_{\mathrm{slow}}}^{+\infty} q(\tau)d\tau = -\kappa \times q(\tau_d>\tau_{\mathrm{slow}})
\end{equation}

And the \gls{efe} becomes:

\begin{equation}
    G(d) \simeq -[\mathrm{EIG}(d) -\kappa q(\tau_d>\tau_{\mathrm{slow}})]
\end{equation}

Tests that are unlikely to last more than $\tau_{\mathrm{slow}}$ are judged equally, based on their epistemic merit alone (since then $G(d)\simeq -\mathrm{EIG}(d)$). However, tests that may exceed this duration will be penalized. $\kappa$ parameterizes the maximum penalty that can be incurred to slow tests. 

\begin{figure}
    \centering
    \includegraphics[width=0.875\linewidth]{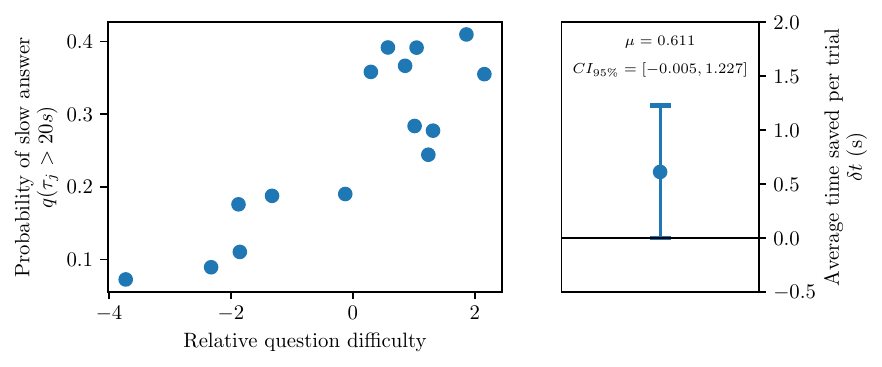}
    \caption{\textbf{Simulation of the penalization of slow tests}. The left plot represents the probability that a participant takes more than 20 seconds to answer a test as a function of the question difficulty. The right plot represents an estimate of the time saved per trial by factoring the expected utility in the selection of questions (black bars indicate the 95\% credible interval of the mean).}
    \label{fig:durations}
\end{figure}

We tested this strategy as a minor modification of our computerized adaptive testing experiment (\S\ref{section:adaptive_testing}).
One important difference is the need to track the probability distribution of $\tau_d$ for each test $d$. We start by assuming that $\tau_d$ follows a log-normal distribution with parameters $\mu_d$ and $\eta_d$.  We then learn $q(\mu_d)$ and $q(\eta_d)$ together with the other relevant parameters ($\theta$ and $\delta$) using the same variational inference approach as before. The expected utility term in the \gls{efe} is estimated by sampling from $q$, as follows:

\begin{equation}
    \mathbb{E}_{q(\tau_d)}[\log p^{\ast}(\tau_d)] = -\kappa  \int_{\tau_{\mathrm{slow}}}^{+\infty} q(\tau)d\tau \simeq -\dfrac{\kappa}{S} \sum_{s=1}^S \mathbbm{1}(\tau_s>\tau_{\mathrm{slow}})
\end{equation}

\noindent where $\tau_s\sim \text{LogNormal}(\mu_{ds},\eta_{ds})$, $\mu_{ds}\sim q(\mu_{d})$, and  $\eta_{ds}\sim q(\eta_{d})$. For purposes of illustration, we simulated the approach on our oracle (200 participants). This reveals differences in the rate of slow answers between different questions, which seem to be mostly explained by differences in difficulty (Figure \ref{fig:durations}). As a rough estimate of the efficiency of the approach, we computed the time saved at each trial as follows. We determined, at each step of the experiment, the optimal tests prescribed by the \gls{eig} alone ($\hat{d}_{EIG}$) and the \gls{efe} ($\hat{d}_{G}$). Using the oracle data, we calculated the difference $\delta t = \tau_{i,\hat{d}_{EIG}}-\tau_{i,\hat{d}_{G}}$ between the times participant $i$ took to answer these questions. We find that $\delta t$ is positive on average, with a mean $\mu\simeq 0.6 \pm 0.6$ s/trial. While this number does not provide a truly counterfactual measure, it gives an order of magnitude for the potential time savings and suggests that the method may prove beneficial when the variance in the tests' response times is even higher. The gain appears limited in the present experiment, presumably because all tasks are relatively similar. 

\section{\label{appendix:active_inference_mab}A more systematic evaluation of active inference in experimental bandit settings}

We compare the performance of active inference and traditional bandit algorithms for adaptive experimentation in two setups (A and B). In both setups, each treatment $1\leq d \leq k$ is an arm in a multi-armed bandit, with an unknown latent reward $\theta_d$. In the setup A, the unknown reward of $d$ is $\theta_d \in [0,1]$. The administration of treatment $d$ produces an observation $y\sim \text{Bernoulli}(\theta_d)$. In the setup B, the unknown reward is $\theta_d=\frac{1}{2}(\theta_{d,z=0}+\theta_{d,z=1})$ where $z \in \{0,1\}$ denotes two different groups of participants. The administration of $d$ produces an observation $y \sim \text{Bernoulli}(\theta_{d,z})$ where $z$ is the group to which the current participant belongs. Setup B is more difficult because each observation produces less information about $\theta_d$ on average. In addition, the information gained by delivering a treatment depends on both $d$ and $z$; this means that exploration strategies that factor this dependence may perform better.

We simulate active inference, Thompson sampling, and exploration sampling in both scenarios while varying the number of treatments $k\in\{5,10,30\}$. For every configuration and algorithm, we run 300 simulations. In each simulation, we draw random rewards $\theta\sim\text{Beta}(2,2)$. We evaluate each algorithm on two performance measures: the policy regret (which measures the opportunity cost of retaining the estimated optimal treatment after the end of the experiment, \citealt{kasy2021adaptive}) and the ``average'' per-trial regret that measures the opportunity costs accumulated throughout the experiment. Exploration sampling is designed to minimize the policy regret, and Thompson sampling is known to minimize the average regret.

\begin{figure}[p]
    \centering
    \includegraphics[width=0.95\linewidth]{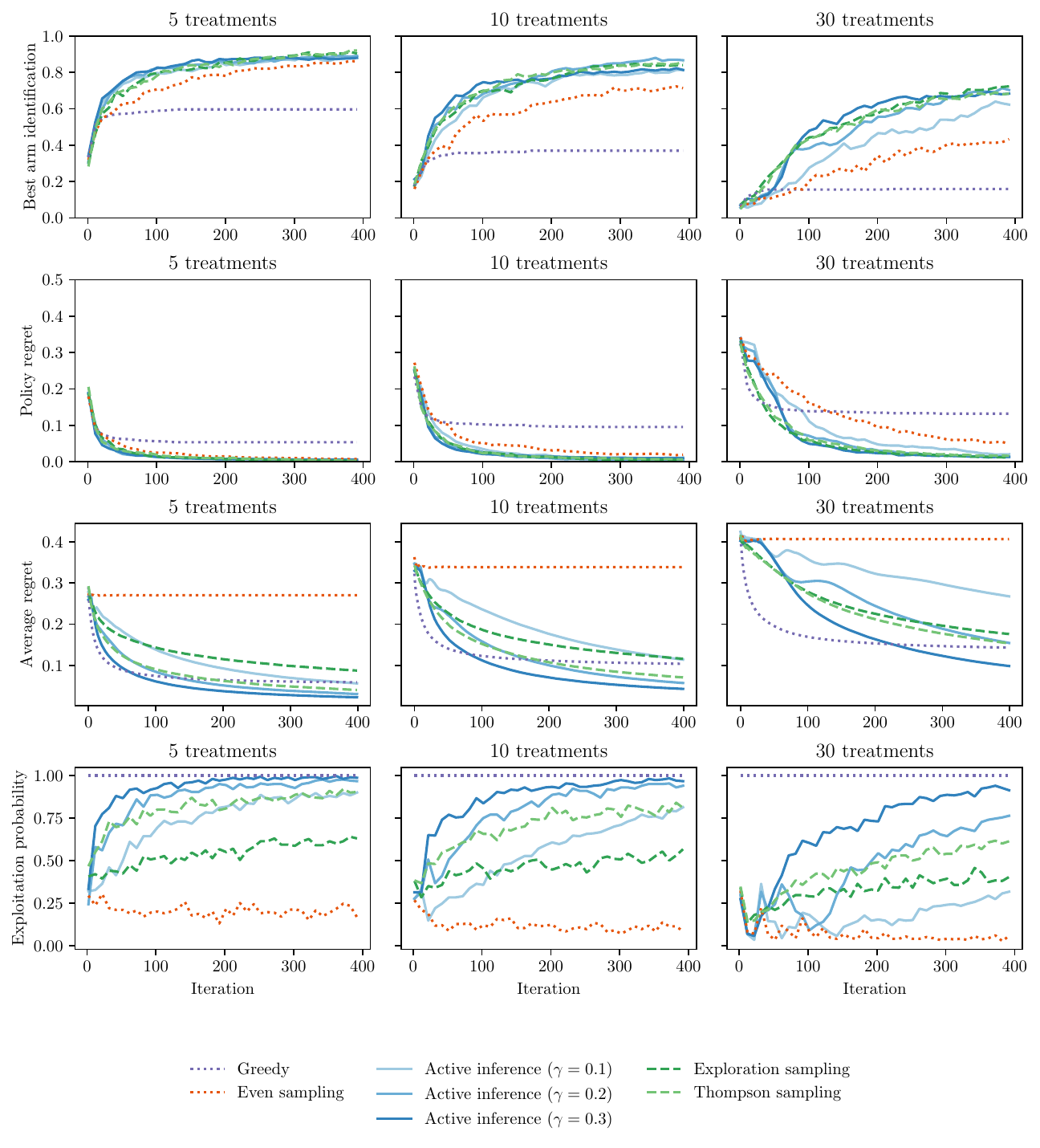}
    \caption{\textbf{Comparison of active inference and traditional bandit algorithms for adaptive experimentation (setup A)}. Performance is evaluated in terms of the best arm identification rate (top plots; higher is better), the policy regret (second row; lower is better) and average regret (third row; lower is better). Differences in behavior are measured in terms of the exploitation probability (bottom plots). \textbf{Best arm identification rate} is the probability that the estimated optimal treatment is the true best treatment. \textbf{Policy regret} is the difference between the true expected utility of the truly optimal treatment, and the true expected utility of the treatment believed to be optimal at time $T$.  \textbf{Average regret} is evaluated as the difference between the average reward at time $T$ if the best treatment had always been chosen (i.e. $\max_d \mathbb{E}[r_d]_{\text{true}}$), and the actual average reward (i.e. $\frac{1}{T}\sum_{t=1}^T \mathbb{E}[r_t]_{\text{true}}$). The \textbf{exploitation probability} measures the rate at which each algorithm administers the treatment with the highest estimated utility. 
    \label{fig:performance_systematic_simple}}
\end{figure}

\begin{figure}[p]
    \centering
    \includegraphics[width=0.95\linewidth]{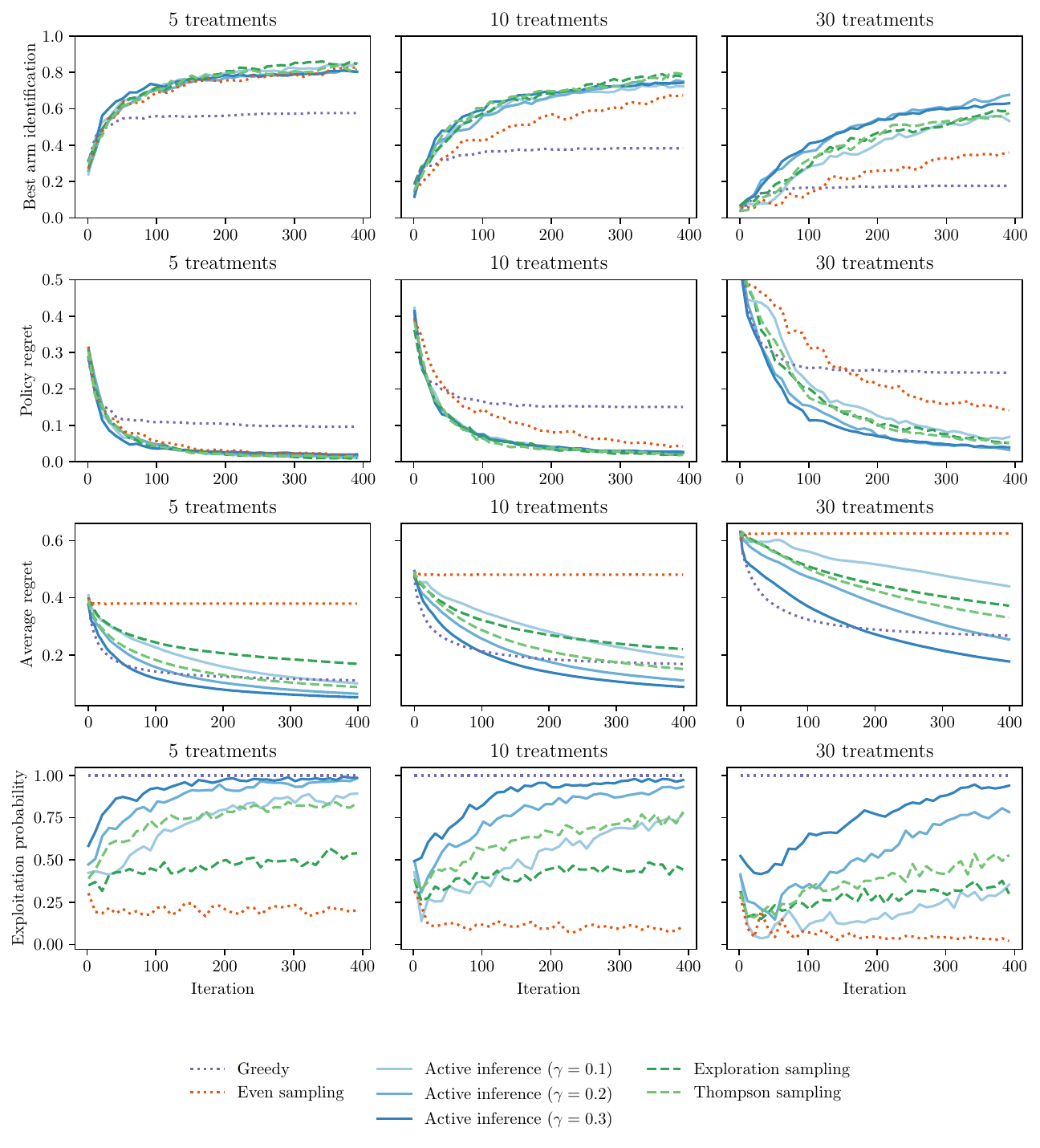}
    \caption{\textbf{Comparison of active inference and traditional bandit algorithms for adaptive experimentation (setup B)}. Performance is evaluated in terms of the best arm identification rate (top plots; higher is better), the policy regret (second row; lower is better) and average regret (third row; lower is better). Differences in behavior are measured in terms of the exploitation probability (bottom plots).
    \label{fig:performance_systematic}}
\end{figure}

Figures \ref{fig:performance_systematic_simple}  and \ref{fig:performance_systematic} show the results for the scenarios A and B, respectively. In scenario A, with appropriate priors (parameterized by $\gamma$, cf. \ref{section:adaptive_treatment}), active inference achieves average regrets comparable to Thompson sampling. However, policy regret decreases more slowly (as a function of the number of observations) when the treatment-space is large. In scenario B, with appropriate choice of priors, active inference outperforms Thompson and exploration sampling on both metrics. A plausible explanation is that the \gls{eig} term in the \gls{efe} produces more efficient exploration, factoring the dependency between $\theta_k$ and $z$: active inference indeed incorporates contextual information when deciding which action would provide more information gain. In other words, the balance of exploration and exploitation in the \gls{efe} produces a behavior that tends to promote the minimization of the policy and average regret, even when the optimization procedure is myopic (i.e., not considering the future after the next step). This makes active inference an interesting alternative to the two reference algorithms. This is especially true when ``smart'' exploration yields superior information gains. 

\section{\label{appendix:choosing_gamma}Constraining $\gamma$ in multi-armed bandit setups}

The expected free energy weighs epistemic and pragmatic costs, and the balance between the two is fixed by the choice of the prior specifying the pragmatic component \eqref{eq:efe_simple}. Often, the hyperparameters of the prior express degrees of freedom that give direct control over the tradeoff. For instance, in the example introduced in \S\ref{section:adaptive_treatment}, the hyperparameter $\gamma$ scales the expected utility term, thereby dictating the balance between exploration and exploitation. The question for the researcher is then, how to choose an appropriate value for $\gamma$? 

\begin{figure}
    \centering
    \includegraphics[width=0.67\linewidth]{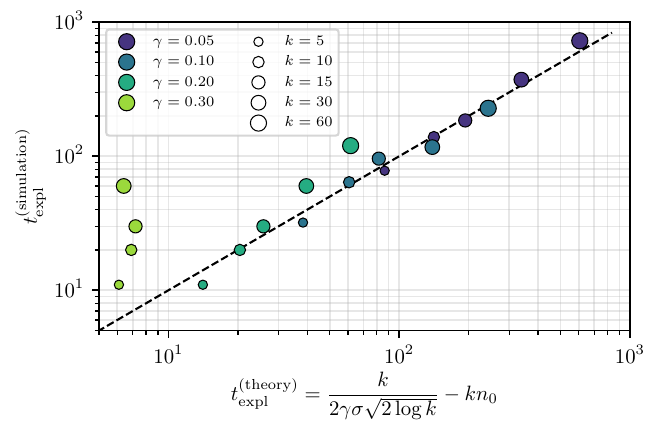}
    \caption{\textbf{Duration of the exploration phase in active inference over a Beta-Bernoulli multi-armed bandit}. We compare the duration of the exploration phase in simulations for different values of $\gamma$ and $k$ with the duration predicted by our qualitative formula. In simulations, the duration of the exploration is the time from which the number of treatments effectively explored over the last $k$ iterations drops below $k/2$. The qualitative prediction no longer holds if $\gamma\gtrsim\gamma_{\text{greedy}}$.}
    \label{fig:exploration_time}
\end{figure}

Let us assume a Beta Bernoulli multi-armed bandit setup, featuring $k$ treatments with utilities $(u_d)_{1\leq d \leq k}$ drawn in $[0,1]$, and binary observations $y\sim Bernoulli(u_d)$. Let $\alpha_d$ and $\beta_d$ be the pseudo-counts of successes and failures for each treatment, and $n_d\equiv\alpha_d+\beta_d$, such that $\mathbb{E}[u_d{\mid}\alpha_d,\beta_d]=\alpha_d/n_d$. When $n_d$ is large, the EIG \eqref{eq:analytical_eig} is approximately $\sim 1/(2n_d)$ \citep{Markovi2021}. This simplified expression emphasizes the diminishing returns of information acquisition over time, whose contribution to the objective function is progressively eclipsed by the $\gamma$-weighted utility term of the \gls{efe} in the long run. As a result, a growing number of treatments will be discarded during the experiment, at a pace determined by $\gamma$. Thus, the choice of $\gamma$ determines the ``duration'' of the exploration phase in the experiment. To see how, let us assume a steady state, where the \gls{efe} gives the following equation for the best arm $\hat{d}$ and
any other arm $d$:
\begin{equation}
    \frac{1}{2n_{\hat{d}}} + \gamma u_{\hat{d}} = \frac{1}{2n_d} + \gamma u_d
    \quad\Longrightarrow\quad
    \frac{1}{2n_d} - \frac{1}{2n_{\hat{d}}} = \gamma(u_{\hat{d}}- u_d)
\end{equation}
\noindent where $u_d$ is, for simplicity, the actual (but in fact unknown) utility of treatment $d$. As $t\to\infty$, $n_{\hat{d}}\to\infty$, and arm $d$ converges to:
\begin{equation}
    n_d \;\xrightarrow{t\to\infty}\; N_{\infty,d} = \frac{1}{2\gamma(u_{\hat{d}}- u_d)}
\end{equation}
In the exploration regime where all arms are visited approximately equally, we have $n_d-n_0 \simeq t/k$, where $t$ is the number of iterations and $k$ the total number of treatments. Plugging this expression into the above equation and solving for $t$ gives:
\begin{equation}
    t_d^{\mathrm{stop}} + kn_0 \simeq \frac{k}{2\gamma(u_{\hat{d}}- u_d)}
\end{equation}
Arms are thus abandoned in order of decreasing utility gap. In simulations, however, the duration of the exploration phase is not defined by the last time the second-best arm is explored, but by the iteration at which the bulk of arms stops being explored. We therefore approximate the relevant utility gap by the gap between the best arm and a typical arm. For utilities drawn from a common distribution with standard deviation $\sigma$, and assuming approximate normality, this gap scales as
$\sigma\sqrt{2\ln k}$, giving:
\begin{equation}
    \label{eq:tc}
    t_{\mathrm{expl}} \simeq \frac{k}{2\gamma\sigma\sqrt{2\log k}} - kn_0
\end{equation}
where $n_0=\alpha_0+\beta_0=2$, reflecting the prior. Thus, as expected, larger values of $\gamma$ are associated with shorter exploration times. Although highly idealized, our derivation provides an appropriate description for a broad range of parameters (Figure \ref{fig:exploration_time}).

We can now propose qualitative bounds on $\gamma$. First, for utility-driven optimization to occur, the exploration phase must be shorter than the total number of iterations in the experiment, giving:

\begin{equation}
\gamma_{\mathrm{expl}} \simeq \frac{k}{2(T + 2k)\sigma\sqrt{2\log k}}
\end{equation}

Below this bound, active inference will assign these treatments almost evenly during the experiment. At the other extreme, \eqref{eq:tc} indicates a critical value $\gamma_{\mathrm{greedy}}$ above which there is no exploration phase at all, thus approaching greedy maximization:

\begin{equation}
    \gamma_{\mathrm{greedy}} \simeq \dfrac{1}{4\sigma\sqrt{2\log k}}
\end{equation}

Additionally, requiring that the exploration phase lasts long enough to reduce the uncertainty in the treatments' utilities below the utility standard deviation $\sigma$ gives:

\begin{equation}
    1/\sqrt{n_d} \lesssim \sigma \Rightarrow \gamma \lesssim \sigma
\end{equation}

While these bounds are not predicated on any particular performance criterion (such as cumulative regret, policy regret, or best-arm identification), they provide indicative constraints informed by key parameters of the experimental setup.

\section{\label{appendix:performance}Online performance}

\begin{figure}[H]
    \centering
    \centering
    \begin{subfigure}[t]{0.475\textwidth}
    \centering
    \includegraphics[width=1\linewidth]{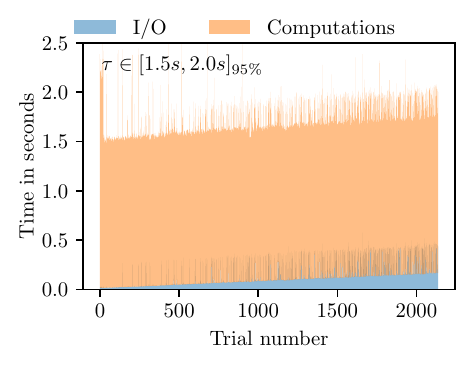}
    \caption{\textbf{Experiment 1}.}
    \label{fig:performance_test}
    \end{subfigure}\hfill\begin{subfigure}[t]{0.475\textwidth}
    \centering
    \includegraphics[width=1\linewidth]{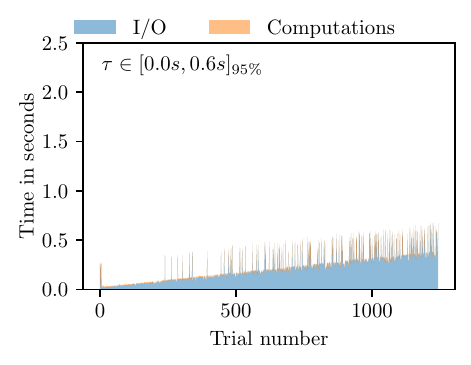}
    \caption{\textbf{Experiment 2}.}
    \label{fig:performance_treatment}
    \end{subfigure}
    \caption{\textbf{Performance of the adaptive design in Experiments 1 and 2}. Performance is measured via the time taken to deliver each trial, in seconds. This time is divided into ``I/O'' (the retrieval of the data from the database) and the computations themselves (which involve the derivation of the posterior distribution and the objective to optimize).
    \label{fig:performance}}
\end{figure}

We deployed the experiment on an AWS EC2 instance with 16 vCPU and 32 GB of memory, accepting up to two participants at a time. The experiment was simulated with 200 bots on the same machine before recruiting participants, in order to verify that performance would be sufficient\footnote{PsyNet makes it easy to test an experiment directly on the server where it will be eventually deployed. For instance, the following command tests an experiment by recruiting three bots in parallel: \texttt{psynet ssh --app my-experiment test --n-bots 3 --parallel}}. In the real experiment, trials were delivered reasonably fast, under two seconds per trial in the first experiment and half a second per trial in the second (see Figure \ref{fig:performance}). 

\end{document}